\documentclass[useAMS,usenatbib,twocolumn]{aa}
\usepackage{graphicx}
\usepackage{amsmath}
\usepackage[scr=esstix]{mathalpha} \usepackage{amssymb}
\usepackage{cases}
\usepackage{empheq}
\usepackage{xspace}
\usepackage{commath}
\usepackage{bm} 
\usepackage{balance}
\usepackage{mathtools}
\usepackage{stmaryrd}
\usepackage{tabulary}
\usepackage{enumitem}

\usepackage{txfonts}
 \usepackage{orcidlink}
 \usepackage{printlen} \usepackage{subcaption}
\renewcommand{\thesubfigure}{\alph{subfigure}}

\usepackage{CJK}

\usepackage{booktabs}
\usepackage[flushleft]{threeparttable} \usepackage{beramono}

\defcitealias{BinneyLacey1988}{BL88}
\defcitealias{BinneyTremaine2008}{BT08}

\usepackage{acronym}

\newacro{DM} {dark matter}
\newcommand{\DM}{\ac{DM}}

\newacro{AGN}{active galactic nuclei}
\newcommand{\AGN}{\ac{AGN}}

\newacro{SB}{superbubble}
\newacroplural{SB}[SBs]{superbubbles}
\newcommand{\SB}{\ac{SB}}
\newcommand{\SBs}{\acp{SB}}

\newacro{NFW}{Navarro--Frenk--White}
\newcommand{\NFW}{\ac{NFW}}

\newacro{DF}{distribution function}
\newacroplural{DF}[DFs]{distribution functions}
\newcommand{\DF}{\ac{DF}}
\newcommand{\DFs}{\acp{DF}}

\newacro{FP}{Fokker--Planck}
\newcommand{\FP}{\ac{FP}}

\newacro{MW}{Milky Way}
\newcommand{\MW}{\ac{MW}}

\newacro{GMC}{Giant Molecular Cloud}
\newacroplural{GMC}[GMCs]{Giant Molecular Clouds}

\newcommand{\GMCs}{\acp{GMC}}

\newacro{SFR}{star formation rate}
\newacroplural{SFR}[SFRs]{star formation rates}
\newcommand{\SFR}{\ac{SFR}}
\newcommand{\SFRs}{\acp{SFR}}
\newcommand{\Mbh}{\text{M}_{\bullet}} \newcommand{\Msol}{\text{M}_{\odot}}  \newcommand{\kpc}{\text{kpc}} \newcommand{\pc}{\text{pc}}  \newcommand{\Myr}{\text{Myr}} \newcommand{\Gyr}{\text{Gyr}} \newcommand{\kmPs}{\text{km/s}} \newcommand{\erg}{\text{erg}} \newcommand{\ru}{\mathrm{u}} \newcommand{\re}{\mathrm{e}} \newcommand{\ri}{\mathrm{i}} \newcommand{\dirac}{ \delta_{\mathrm{D}}} \newcommand{\dd}{\mathrm{d}} \newcommand{\p}{\partial} \newcommand{\bk}{\boldsymbol{k}}  \newcommand{\br}{\boldsymbol{r}} \newcommand{\bv}{\boldsymbol{v}} \newcommand{\bL}{\boldsymbol{L}} \newcommand{\bJ}{\boldsymbol{J}} \newcommand{\btheta}{\boldsymbol{\theta}} \newcommand{\uv}[1]{\hat{\boldsymbol{\re}}_{#1} } \newcommand{\uvu}[1]{\hat{\boldsymbol{\ru}}_{#1}} \newcommand{\scrf}{\mathscr{f}} 
\newcommand{\scrr}{\mathscr{r}} 
\newcommand{\scrrc}{\mathscr{r}_{\text{c}}} 
 
\newcommand{\calP}{\mathcal{P}} 
 
\newcommand{\calE}{\mathcal{E}} 
 
\newcommand{\calO}{\mathcal{O}} 
\newcommand{\calEmax}{\mathcal{E}_{\text{max}}} 
 
\newcommand{\Lc}{L_{\text{c}}} 
\newcommand{\rc}{r_{\text{c}}} 
\newcommand{\vc}{v_{\text{c}}}

\newcommand{\Dz}{D_{0}} 
\newcommand{\DEEmax}{D_{EE}^{\text{max}}}
  
\newcommand{\kmax}{k_{\text{max}}} 
\newcommand{\kparl}{k_{\parallel}} 
\newcommand{\kperp}{k_{\perp}}

\newcommand{\fiso}{f^{\text{iso}}} 
 
\newcommand{\giso}{g^{\text{iso}}} 
 
\newcommand{\Mvir}{M_{200}} 
\newcommand{\Rvir}{R_{200}} 
\newcommand{\rhocrit}{\rho_{\text{crit}}} 
\newcommand{\cNFW}{c_{\text{NFW}}} 
\newcommand{\DEErad}{D_{EE}^{\, \text{rad}}} \newcommand{\trel}{\tau_{\textrm{relax}}} \newcommand{\tdiff}{\tau_{\textrm{D}}} \newcommand{\tbal}{\tau_{\textrm{bal}}} \newcommand{\thalo}{T_{\textrm{halo}}} \newcommand{\tdyn}{\tau_{\text{dyn}}} \newcommand{\rmax}{r_{\textrm{max}}} \newcommand{\ra}{r_{\textrm{a}}} \newcommand{\rp}{r_{\textrm{p}}} \newcommand{\vr}{v_{r}} \newcommand{\vt}{v_{t}}   \newcommand{\vparl}{v_{\parallel}} \newcommand{\vperp}{v_{\perp}} \newcommand{\thetar}{\theta_{r}} \newcommand{\thetatheta}{\theta_{\theta}} \newcommand{\thetaphi}{\theta_{\phi}}  \newcommand{\Omegar}{\Omega_{r}} \newcommand{\Omegaphi}{\Omega_{\phi}} \newcommand{\rhohalo}{\rho_{\text{halo}}} \newcommand{\Vhalo}{V_{\text{halo}}} \newcommand{\Rhalo}{R_{\text{halo}}} \newcommand{\Lhalo}{L_{\text{halo}}} \newcommand{\Rdisc}{R_{\text{disc}}} \newcommand{\Hdisc}{H_{\text{disc}}} \newcommand{\Mhalo}{M_{\text{halo}}} \newcommand{\Rbub}{R_{\text{bub}}} \newcommand{\Rbubin}{R_{\text{in}}} \newcommand{\Rbubout}{R_{\text{out}}} \newcommand{\Tbub}{T_{\text{bub}}} \newcommand{\rhobub}{\rho_{\text{bub}}} \newcommand{\rhobubin}{\rho_{\text{in}}} \newcommand{\rhobubout}{\rho_{\text{out}}}   \newcommand{\Nbub}{N_{\text{bub}}}  \newcommand{\dotnbubz}{\dot{n}_{\text{bub}, 0}}  \newcommand{\dotnbub}{\dot{n}_{\text{bub}}}  \newcommand{\dotNbub}{\dot{N}_{\text{bub}}}  \newcommand{\dotNeff}{\dot{N}_{\text{eff}}}  \newcommand{\Mbub}{M_{\text{bub}}} \newcommand{\Ebub}{E_{\text{bub}}} \newcommand{\Sigmabub}{\Sigma_{\text{bub}}} \newcommand{\Mkv}{Markovian} \newcommand{\nMkv}{non-Markovian} \newcommand{\Tsim}{T_{\textrm{sim}}} \usepackage{soul} \usepackage{xcolor}
\definecolor{aquamarine}{rgb}{0.5, 1.0, 0.83}
\definecolor{cyan}{rgb}{0.8784,1.0,1.0}

\usepackage{hyperref}
\hypersetup{colorlinks=true, linkcolor=blue, citecolor=blue, filecolor=blue, urlcolor=blue}

\begin{document}
\begin{CJK*}{UTF8}{mj}
	\title{Coupling baryons to galactic secular dynamics:\\
 I/ Baryonic feedback-driven orbital diffusion of dark matter cusps}
   	\titlerunning{Bubble-driven orbital diffusion of cusps}
   
 	\authorrunning{E. Ko, C. Pichon, K. Tep, J.-B. Fouvry, S. Prunet}
 
  	\author{
  	Eunhee~Ko~(고은희)~\inst{1}\orcidlink{0000-0002-8130-8044}\and
        Christophe~Pichon\inst{1,2}\thanks{Corresponding author: \href{mailto:pichon@iap.fr}{pichon@iap.fr}}  \orcidlink{0000-0003-0695-6735}\and
 	Kerwann~Tep\inst{3}\orcidlink{0009-0002-8012-4048} \and
        Jean-Baptiste~Fouvry\inst{1}\orcidlink{0000-0002-0030-371X} \and
    	Simon~Prunet~\inst{4}\orcidlink{0000-0002-1755-4582}
                   }       
 	\institute{
            Institut d'Astrophysique de Paris, CNRS and Sorbonne Universit\'e, UMR 7095, 98 bis Boulevard Arago, F-75014 Paris, France
        \and
Kyung Hee University, Dept. of Astronomy \& Space Science, Yongin-shi, Gyeonggi-do 17104, Republic of Korea
	\and
             Observatoire Astronomique de Strasbourg, CNRS UMR 7550, 11 rue de l’Universit\'e, F-67000 Strasbourg, France 
         \and
             Universit\'e C\^ote d’Azur, Observatoire de la C\^ote d’Azur, CNRS, Laboratoire Lagrange, Bd de l’Observatoire, 06304 France
             }
 
\abstract{
To explain how baryonic feedback flattens central dark matter cusps into cores, we develop an analytical kinetic framework for orbital diffusion driven by recurrent supernova explosions. In the limit of weak deflections induced by local potential fluctuations, the rate of orbital diffusion of dark matter or stellar particles can be computed as a three-step process involving (i)~quantifying the effect of an individual deflection, (ii)~characterising the statistical impact of many deflections via a local power spectrum, and (iii)~finally phase-averaging this local diffusion over their orbits to compute global diffusion coefficients as a function of orbital parameters. 
This \emph{orbit-averaged Fokker-Planck} strategy is applied to the cusp-to-core transformation of dark matter profiles induced by supernova explosions, represented as superbubbles. Simple analytic expressions for the corresponding relaxation rates are presented as combinations of characteristic size, lifetime, and density of a superbubble for a given orbit. 
They are successfully validated against Monte Carlo simulations. 
The initial rate of cusp flattening is derived from the corresponding Fokker-Planck equation, predicting ${\sim \! 2.0 \, \Gyr}$ for dwarf galaxies, and ${\sim \!10.5 \, \Gyr}$ for Milky Way-like galaxies. We further quantify how finite spatial resolution truncates the fluctuation spectrum and underestimates diffusion rates in numerical simulations.
The methodology could be straightforwardly applied to other physical processes such as the heating of self-gravitating stellar discs or the dissolution of galactic bars by giant molecular clouds and supernovae. 
}

\maketitle

\section{Introduction}\label{sec:Introduction}
\subsection{Astrophysical context}\label{subsec:Astro}
The structure of \DM\@ haloes represents a useful test for modern cosmological paradigms and theories of galaxy formation. While collisionless \DM\@ simulations robustly predict steep inner density profiles \citep[e.g.,][]{DubinskiCarlberg1991,Navarro1996,Navarro1997,Fukushige1997,Moore1999}, observations often display flatter cores in the centres of galaxies \citep[e.g.,][]{Burkert1995,Oh2008,deBlok2010,DelPopolo2021}. A well-established candidate for resolving this tension invokes baryonic physics. The energetic feedback from supernovae, which drives the expansion of large superbubbles in the interstellar medium, induces rapid, localised fluctuations in the galactic gravitational potential. These stochastic perturbations cumulatively transfer energy to the collisionless \DM\@ particles, dynamically heating them and causing their orbits to migrate, thereby possibly transforming an initial cusp into a core.

Numerical and theoretical efforts have explored this baryonic feedback mechanism to understand the expected structural evolution of cusps \citep[e.g.,][]{Pontzen2012,Peirani2017,Yadav2017,Lazar2020,Kipper2025}. The effect of such secular processes is not isolated: it impacts the  galactic ecosystem, linking the behaviour of giant molecular clouds and stellar feedback to the underlying \DM\@ dynamics \citep[e.g.,][]{Licquia2015,BlandHawthorn2016,SylosLabini2023}.

To characterise the dynamical heating driven by such weak, incoherent perturbations over secular timescales, one must appeal to kinetic theory, specifically the \FP\@ equation. The foundation of this approach in astrophysics stems from  work on stellar encounters, diffusion coefficients, and relaxation 
\citep{Chandrasekhar1943,Cohen1950,SptizerSchwarzschild1951,Rosenbluth1957}. It was  advanced through  studies of stellar  core collapse 
\citep{Peebles1972,BahcallWolf1976,FrankRees1976,CohnKulsrud1978,Cohn1979,Kandrup1980,Cohn1980}
and also   shaped by  statistical  and plasma physics 
\citep{PicardWillis1974,PicardWillis1977,Brizard2004,Brizard2009}.

In galactic dynamics, extending a local description to encompass the global orbital structure of a galaxy relies on the orbit-averaging framework exemplified by \citet[hereafter \citetalias{BinneyLacey1988}]{BinneyLacey1988}. For systems governed by a dominant, smooth mean-field potential, the secular evolution of the \DF\@ can be mapped directly onto the space of integrals of motion \citep{Dejonghe1986,Hernquist1990,Widrow2000,Evans2006,BinneyTremaine2008}.
Over the past few decades, these secular diffusion models have been extensively applied, and have seen a renewed surge of interest
across environments ranging from dense galactic nuclei \citep{EvansCollett1997,Bar-Or2016,Vasiliev2017} to extended galactic discs and haloes \citep{Lacey1984,ChernoffWeinberg1990,Fouvry2021,Tep2022,Lane2022,Petersen2024,Modak2026Interstellar}.

In this paper, we aim to present a versatile strategy to model the orbital diffusion of \DM\@ cusps driven by repeated supernova explosions. Our approach computes the diffusion rate in three steps: first, we quantify the localised effect of a single velocity deflection when entering a \SB\@; second, we characterize the collective impact of multiple deflections through the corresponding local power spectrum; and third, we phase-average the corresponding local diffusion coefficients over particle orbits to obtain global diffusion coefficients as functions of the orbital parameters.

To explain cusp-to-core transformations, numerous studies have investigated how time-varying potential fluctuations driven by baryonic processes reshape central \DM\@ halo profiles. These investigations have adopted various idealizations to characterize feedback-driven perturbers. For example, \citet{Pontzen2012} modeled feedback-induced fluctuations using a time-varying harmonic oscillator that recurrently transfers energy to the \DM\@ halo. \citet{Ogiya2014} focused on resonant coupling between perturbation waves and \DM\@ orbits, highlighting that energy transfer is maximized when their characteristic timescales coincide. Similarly, \citet{ElZant2016} examined the statistical properties of stochastic, homogeneous gas fluctuations. More recently, refined observational data have enabled a more realistic characterization of \SB\@ physical properties and their associated dynamical friction \citep{Kipper2025}, while hydrodynamical simulations have quantified the fluctuation power spectra of the interstellar medium \citep{Modak2026Characterizing}.

Although these earlier approaches have highlighted key aspects of the underlying dynamics, a comprehensive treatment must simultaneously incorporate three essential ingredients: (i)~the detailed physical properties of individual perturbing sources, (ii)~their inhomogeneous spatial distribution, and (iii)~the resulting secular evolution of the phase-space \DF\@ as a function of orbital invariants. Constructing such an integrative theoretical framework is the primary objective of this work.

By revisiting the orbit-averaged \FP\@ equation based on \citetalias{BinneyLacey1988}, we develop a theory that connects observationally motivated local fluctuation power spectra, constructed directly in physical space, to global orbital diffusion coefficients. This formalism avoids the need to explicitly construct action-angle coordinates to evaluate the associated Jacobian, which can be difficult to obtain in realistic potentials. Moreover, it can take into account global inhomogeneities in the distribution of perturbing sources, extending the applicability of Chandrasekhar's original framework beyond its idealized assumptions.
As a result, we derive simple analytic expressions for the resulting relaxation rates. These expressions are formulated for purely radial and general orbits of \DM\@ particles. We successfully validate these analytical coefficients against dedicated Monte Carlo simulations, allowing us to derive the initial rate of cusp flattening as a function of the \SB\@ physics. Other applications will be presented in the companion paper.

\subsection{Outline of the paper}
The paper is structured as follows. Section~\ref{sec:Theory} establishes the theoretical framework, detailing the local velocity deflection, the formulation of power spectra, and the global phase-averaging required to calculate the orbit-averaged diffusion coefficients. Section~\ref{sec:Results} presents the resulting analytical diffusion coefficients and their approximate scaling relations for various orbital families.  Section~\ref{sec:NumericalValidation}\@  validates these results through both \Mkv\@ and \nMkv\@ numerical simulations. Section~\ref{sec:CuspCoreTransformation} applies this formalism directly to the cusp-to-core transformation, deriving the inner density slopes for different \DFs\@ and calculating relaxation timescales for Milky Way-like, dwarf, and giant galaxies. Finally,  Section~\ref{sec:Discussion} discusses the fundamental limitations of our methodology, and Section~\ref{sec:ConclusionsAndPerspectives} concludes with perspectives on applying this framework to broader contexts like disc heating, bar dissolution, and constraining \DM\@ models. Throughout the paper, technical details are provided in the Appendices, with relevant sections referenced in the main text where appropriate.

\section{Theory: Orbit-averaged diffusion coefficients} \label{sec:Theory}
\begin{figure*}[htbp]
	\center\includegraphics[width=0.80\textwidth]{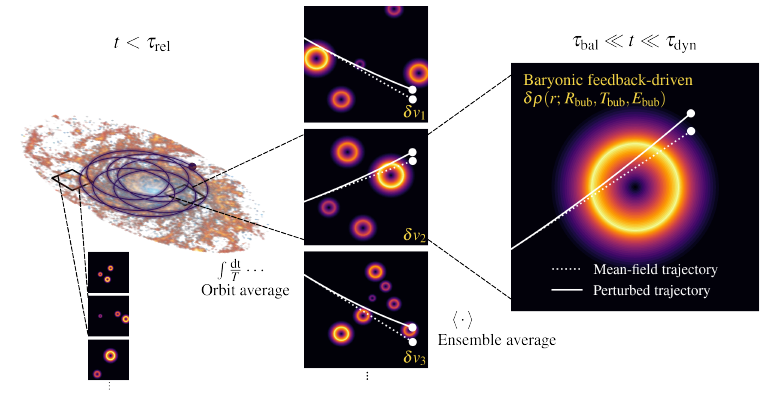}
	\caption{
	An overview of the multi-scale procedure to derive the orbit-averaged diffusion coefficients.\protect\footnotemark\, To gauge the long-term diffusion in invariants (e.g., energy $E$, angular momentum $L$) induced by a series of perturbing sources, the orbit-averaging framework approximates the global diffusion as the additive sum of local diffusion contributions, assuming correlations across different scales can be neglected. The procedure therefore follows three steps: (i)~compute deflection of a locally straight orbit by a single \SB\@ (right panel), (ii)~perform ensemble average of local deflections over different realizations of homogeneously distributed \SBs\@ and orbital phases (middle panel), and (iii)~carry out global phase-averaging along the orbit in the inhomogeneous medium (left panel).
	}
	\label{fig:Schematic}
\end{figure*}
To characterize the dynamical heating induced by weak, incoherent perturbations over the secular timescale, we revisit the orbit-averaged \FP\@ equation. We consider a \DM\@ particle of velocity $\bv$ at position $\br$, orbiting in a spherical system (see \appref{app:sec:FrameAndCoordinates} for the relevant coordinate systems). We assume the system is in a quasi-stationary state: it is characterized by a smooth potential $\Phi_{0}(r)$ and Hamiltonian ${H_{0} = v^{2}/2 + \Phi_{0}(r)}$, and the corresponding \DF\@ can be written as ${f = f_{0}(\bJ)}$ with $(\bJ, \btheta)$ a set of action-angle variables \citep[hereafter \citetalias{BinneyTremaine2008}]{BinneyTremaine2008}. Due to the assumed spherical symmetry of the \DM\@ halo, the Hamiltonian $H_{0}$ can be re-written  as a function of two invariants, such as $(J_{r}, L)$ or $(E, L)$ for radial action $J_{r}$, angular momentum $L$, and energy $E$. Our goal is to derive the diffusion coefficients in terms of such invariants. We illustrate the schematic derivation in \figref{fig:Schematic}. Let us first list a series of assumptions we make:

\begin{itemize}[label=$\circ$]
	\item \textit{Perturbative approach.} Over the secular timescale, the impact of \SBs\@ is assumed to be stochastic, local, weak, and uncorrelated in time and space, while the mean-field potential of the \DM\@ halo is expected to dominate the dynamics. Within this perturbative regime, the orbital diffusion can therefore be described by decoupling a series of local encounters from the secular relaxation, and accumulating their effects globally. This corresponds to \emph{orbit-averaging} (\citealp{Henon1961,Cohn1979}; \citetalias{BinneyLacey1988}). 
		\item \textit{Local homogeneity and global inhomogeneity.} On timescales sufficiently longer than the ballistic timescale of the perturbations, $\tbal$, but shorter than the galactic dynamical timescale, $\tdyn$,  the \SB\@ distribution is approximated as locally homogeneous, and the \DM\@ trajectory as a straight line. On orbital timescales (longer than $\tdyn$ but shorter than the relaxation timescale $\trel$), however, the \DM\@ particle samples the inhomogeneously distributed \SBs\@ set by the galaxy's mean-field spatial variations \citep{Chandrasekhar1943,BinneyTremaine2008,Chavanis2013}. We assume that a timescale separation holds, which allows us to factor the local diffusion coefficients out into multiplication of a homogeneous term and a position-dependent weight that counts the number of perturbers met along the orbit. 
		\item \textit{Decoherence across different scales.} Due to the weak and local perturbations from \SBs\@, the timescale hierarchy, ${\tbal \ll \tdyn \ll \trel}$, allows the two intermediate regimes to be treated separately under the \Mkv\@ approximation. For ${\tbal \ll T \ll \tdyn}$, the \DM\@ particle's trajectory follows its unperturbed, locally straight line between encounters. For ${\tdyn \ll T \ll \trel}$,  the accumulated deflections are globally integrated along the mean-field trajectory. The resulting evolution is therefore \Mkv\@ at both levels: a deflection at one location does not influence subsequent deflections, but their aggregation is updated globally (\citetalias{BinneyLacey1988}; \citealp{Fouvry2021,HamiltonFouvry2024}). 
		\item \textit{Coplanarity of the \DM\@ and \SB\@ disc.} We assume in the main text that the most relevant \DM\@ orbital plane remains coplanar with the \SB\@ disc, in which the perturbing\footnotetext[1]
{Image source of galaxy \href{https://science.nasa.gov/asset/webb/webb-and-hubbles-views-of-spiral-galaxy-ngc-628/}{NGC 628} (left panel): NASA, ESA, CSA, STScI, and the JWST PHANGS team.} sources reside (see \figref{fig:Schematic}).\footnote{This configuration can be applied to a spherically symmetric distribution of \SBs\@. In this case, the results are independent of the orbital plane orientation.} \DM\@ orbits exploring the spherical halo need not be coplanar (see \figref{fig:Frame}) and may be torqued out of their plane in more realistic halo models \citep{RauchTremaine1996}. In a dynamically hot halo, the most effective encounters occur when the \DM\@ orbit is coplanar with the \SB\@ disc, where relative velocities are smallest (e.g., \S{7.4} in \citetalias{BinneyTremaine2008}). Diffusion perpendicular to the disc is less efficient due to both higher relative velocities \citep{Quinn1986,Penarrubia2002} and the lower encounter rate of \SBs\@ away from the disc \citep{Weinberg1987}. We relax this coplanarity assumption in \appref{app:sec:BeyondCoplanarity} by considering misaligned orbital planes. In that more general configuration, the diffusion depends on the orientation of the angular momentum vector, which therefore modifies the shape of diffusion. 
\end{itemize}
Within the realm of these approximations, let us assume that the mean-field \DF\@ in a homogeneous local patch, $f = f_{0}(\bv)$, can be described by the \FP\@ equation in  Cartesian velocity space \citep{Rosenbluth1957}
\begin{equation}\label{eq:FokkerPlanckInCatesianVelocity}
	\frac{\p f}{\p t} = - \frac{\p}{\p v_{i}} d_{v_{i}} f + \frac{1}{2} \frac{\p^{2}}{\p v_{i} \, \p v_{j}} d_{v_{i} v_{j}} f \, , 
\end{equation}
where the first-order (resp. second-order) diffusion coefficients in velocity component $v_{i}$, $d_{v_{i}}$ (resp. $d_{v_{i} v_{j}}$), are defined as the average first (resp. second) moment of velocity deflection $\delta v_{i}$ over the timescale ${\tbal \ll T}$
\begin{align}\label{eq:LocalDiffusionCoefficientDefinition}
 d_{v_{i}} \equiv \frac{\langle \delta v_{i} \rangle}{T} \,; \quad{}
 d_{v_{i} v_{j}} \equiv \frac{\langle \delta v_{i} \delta v_{j} \rangle}{T} \, .
\end{align}
Here the angle bracket $\langle \cdot \rangle$ denotes the average over different realizations of a set of stochastic perturbers (see \figref{fig:Schematic}). Note that in equation~\eqref{eq:LocalDiffusionCoefficientDefinition}, the first-order diffusion coefficient is a rank-1 tensor (a vector), whereas the second-order diffusion coefficient is a rank-2 tensor (a matrix). The velocity deflection $\delta \bv$ is characterized by the perturber's potential fluctuation (see \secref{subsec:LocalVelocityDeflection}), and its statistical properties are determined by the perturber's power spectra (see also \secref{subsec:LocalAveraging}).

Finally, the global diffusion arises from the incoherent accumulation of these local effects along the orbit while also accounting for an inhomogeneous  distribution of \SBs\@. This orbit-averaging process maps the local diffusion coefficient in velocity, defined in equation~\eqref{eq:LocalDiffusionCoefficientDefinition}, to the \emph{orbit-averaged} diffusion coefficients $D_{q_{i}}$ (resp. $D_{q_{i} q_{j}}$), where $q = \{ E, L \}$ (see \secref{subsec:GlobalPhaseAveraging}). Thus, the corresponding orbit-averaged \FP\@ equation in terms of $q$ reads
\begin{equation}\label{eq:FokkerPlanckInEL}
	\frac{\p f}{\p t} = - \frac{\p}{\p q_{i}} D_{q_{i}} f + \frac{1}{2} \frac{\p^{2}}{\p q_{i} \, \p q_{j}} D_{q_{i} q_{j}} f \, .
\end{equation}
In the following, we summarize the \textit{ab initio} derivation of the diffusion coefficients entering  equation~\eqref{eq:FokkerPlanckInEL}.
\subsection{Local velocity deflection by a single encounter}\label{subsec:LocalVelocityDeflection}

At a given position $\br$, the equation of motion for the velocity perturbation, $\delta \bv$, due to a potential fluctuation by a single perturber, $\delta \Phi$, reads
\begin{equation}\label{eq:EquationOfMotion} \delta \dot{\bv} = - \nabla \delta \Phi \, ,
\end{equation}
where the dot denotes the time derivative. Integrating equation~\eqref{eq:EquationOfMotion} over time $T$, during which the test particle experiences the local kick along the $i^{\mathrm{th}}$ coordinate direction, gives
\begin{equation}\label{eq:DeltaV}
	\delta v_{i} = - \ri \! \int \! \frac{\dd^{3} \bk}{(2\pi)^{3}} \, k_{i} \!\int_{0}^{T} \! \dd t \, \delta \Phi_{\bk}( t) \, \re^{\ri \bk \cdot \br(t) } \, ,
\end{equation}
where we use the Fourier transform in space $\br$ using equation~\eqref{eq:FourierTransform}. Note that equation~\eqref{eq:DeltaV} is obtained from a single realization $\delta \Phi_{\bk}( t)$.

\subsection{Local averaging: Power spectra}\label{subsec:LocalAveraging}
\begin{figure}[!htbp] 
	\center\includegraphics[width=0.49\textwidth]{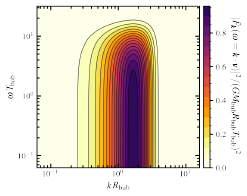}
	\caption{
	Force power spectrum of \SB\@ as a function of spatial frequency $k$, scaled by $\Rbub$, and temporal frequency $\omega$, scaled by $\Tbub$. The peak is located near the inverse of the characteristic \SB\@ parameters. Locally, at fixed velocity $\bv$, the resonance condition ${\omega = \bk \cdot \bv}$ in equation~\eqref{eq:LocalDiffusionCoefficientInVelocity} corresponds to a linear line with slope $\bv$. When orbit-averaging, the change in $\bv(t)$ allows the system to explore different sections of the power spectrum, with their contributions weighted by the inhomogeneity $\Sigmabub$ given by equation~\eqref{eq:SigmabubInR} and the time spent along the trajectory, ${\sim 1/\vr}$.
	}
	\label{fig:PowerSpectrum}
\end{figure}
To quantify the mean impact of perturbations, we adopt a stochastic framework where perturbing sources appear at random positions and times. We assume that within a locally homogeneous patch, the average intensity of local diffusion is proportional to the formation rate of \SBs\@ per unit volume, denoted by $\dotnbub$. Furthermore, we assume that the correlation timescale between consecutive encounters within this local patch is much shorter than the dynamical crossing timescale, such that ${\tbal \ll T \ll \tdyn}$, where $T$ is the timescale during which the test particle experiences the local kick in equation~\eqref{eq:EquationOfMotion}. Averaging the square of equation~\eqref{eq:DeltaV} over distinct realizations of \SB\@ positions and birth times within the local domain yields the second-order local velocity diffusion coefficient:
\begin{equation}\label{eq:LocalDiffusionCoefficientInVelocity}
 d_{v_{i} v_{j}} = \dotnbub \int \! \frac{\dd^{3} \bk}{(2\pi)^{3}} \, k_{i} k_{j} \, \tilde{\calP}_{\bk}(\omega = \bk \cdot \bv) \, ,
\end{equation}
where the local power spectrum is defined as
\begin{equation}\label{eq:LocalPowerSpectrum}
 \, \tilde{\calP}_{\bk}(\omega) \equiv \langle |\delta \tilde{\Phi}_{\bk}(\omega)|^{2} \rangle \, .
\end{equation}
Details on computing the local diffusion coefficients are provided in \appref{app:sec:LocalDiffusionCoefficients}.

Motivated by observations, the potential fluctuation $\delta \Phi$ is initially constructed ab initio from characteristic parameters describing the evolution of \SBs\@ (see \appref{app:sec:ModellingSuperbubbles}). Hence, equation~\eqref{eq:LocalPowerSpectrum} encodes the characteristic spatial and temporal scales of the typical perturbers. An example force power spectrum is illustrated in \figref{fig:PowerSpectrum}. Here, the shape of the power spectrum is determined directly by the spherical geometry of \SBs\@, and its amplitude is maximized at the scale comparable to the perturber's physical size, $\Rbub$, lifespan, $\Tbub$, and mass density, $\rhobub$. Because each \SB\@ only redistributes gas, with zero net mass, the force power spectrum vanishes on large spatial scales, $k \to 0$.

Let us highlight some physics, encoded in equation~\eqref{eq:LocalDiffusionCoefficientInVelocity}. First, an average diffusion occurs only from contributions of the power spectrum that satisfy the resonance condition $\omega = \bk \cdot \bv$: a particle moving at velocity $\bv$ responds to fluctuations whose frequency matches the rate at which it crosses their spatial structure. Therefore, the \SB\@ and the test particle should resonate sharply at specific timescales $1/\omega$ and spatial scales $1/k$. Second, the proportionality, $\dotnbub$, can be understood as a measure for the local \SFR\@ \citep[e.g., see][]{Kipper2025}.

Using the physical relations relating energy, angular momentum, and velocities at fixed position, we can re-express the diffusion coefficient in terms of the invariant energy $E$ and angular momentum $L$ (see \appref{app:subsec:LocalDeflectionInEL}). These are appropriate to characterize the long-term evolution of the system.

\subsection{Global phase averaging}\label{subsec:GlobalPhaseAveraging}

For a local first-order (resp. second-order) diffusion coefficient ${d_{q_{i}}}$ (resp. ${d_{q_{i} q_{j}}}$) in some quantity ${q_{i} \in \{E, L\}}$, the orbit-averaged diffusion coefficient $D_{q_{i}}$ (resp. ${D_{q_{i} q_{j}}}$) is defined as the average over $T \ll \trel$
\begin{align}\label{eq:OrbitAveragingDefinition}
	D_{q_{i}} \equiv \frac{1}{T} \int_{0}^{T} \! \dd t \, d_{q_{i}}(\bv[t]) \, , \text{ and } \, D_{q_{i} q_{j} } \equiv \frac{1}{T} \int_{0}^{T} \! \dd t \, d_{q_{i} q_{j} }(\bv[t]) \, .
\end{align}
Recalling that the number of perturbers varies at each locus in the galactic system, $\dotnbub$ in equation~\eqref{eq:LocalDiffusionCoefficientInVelocity} is now a function of $\br$, and can be generalized as
\begin{align}\label{eq:dotnbub}
	\dotnbub(\br) = \dotnbubz \, \Sigmabub(\br[t]) \, ,
\end{align}
where $\dotnbubz$ is a constant creation rate  of \SBs\@ at the centre per unit volume and $\Sigmabub$ is a dimensionless distribution weight of \SBs\@. Here, $\Sigmabub$ therefore represents the inhomogeneous distribution of perturbers, which weights the degrees of local diffusion as the \DM\@ orbit probes the galactic disc.

From equation~\eqref{eq:OrbitAveragingDefinition}, we note that orbit-averaging changes $\bv(t)$, which in turn updates the resonance condition in equation~\eqref{eq:LocalDiffusionCoefficientInVelocity}. By sweeping through resonance sections, the global diffusion results from a broadened resonant impact, leveraging local resonances between the distribution of perturbing sources and the geometry of the orbit.

To practically perform orbit-averaging, it is convenient to (i)~find an appropriate set of orbital variables instead of time $t$, and (ii)~express the local diffusion coefficient in terms of those variables. To identify such a set, we exploit ergodicity of the orbital motion on invariant tori (e.g., \S{3.5} in \citetalias{BinneyTremaine2008}): along an orbit the actions $\bJ$ stay  invariant, while the angles advance linearly in time. For an orbit-averaging timescale $T$, which is much longer than the dynamical timescale $\tdyn$ but much shorter than the relaxation timescale $\trel$, the orbit covers the torus uniformly, so that the time average of any phase-space function equals its average over the angles:\begin{equation}\label{eq:Ergodicity}
	\int_{0}^{T} \! \frac{\dd t}{T} \, \leftrightarrow \oint_{[0, 2\pi )^{n}} \! \frac{\dd^{n} \btheta}{(2\pi)^{n}} \,  \, ,
\end{equation}
where ${\btheta = (\theta_{1}, \ldots, \theta_{n})}$ are the angle coordinates conjugate to the actions, and $n$ is the number of non-degenerate degrees of freedom. For a generic three-dimensional integrable potential (e.g., a St\"{a}ckel; see \citealt{Tep2025}), ${n = 3}$, whereas in a spherical potential such as that of our \DM\@ halo, each orbit is confined to a fixed plane, so the motion is effectively two-dimensional, giving ${n = 2}$. See \appref{app:sec:OrbitAveraging} for more details.

\section{Results: Orbit-averaged diffusion coefficients}\label{sec:Results}

Let us now present our main results. In \secref{subsec:ModelParameters}, we introduce the parameters adopted in this work: the perturbing source, \SBs\@, a \NFW\@ model galaxy, and diffusion quantities. The prediction of orbit-averaged diffusion coefficients for radial and general (non-radial) orbits is detailed in \secref{subsec:RadialOrbits} and \secref{subsec:GeneralOrbits}, respectively. In \secref{subsec:ScalingRelations}, a simple analytic approximation for the scaling of corresponding diffusion coefficients is presented. 

\subsection{Model quantities}\label{subsec:ModelParameters}

For convenience, we summarize the parameters and quantities used throughout this work in Table~\ref{tab:Parameters}, with each defined in detail in its respective section.
Unless otherwise specified, we adopt the \NFW\@ halo model, whose corresponding potential and density profile are set by the scale radius $\Rhalo$, circular velocity $\Vhalo$, and mass density $\rhohalo$ (see \appref{app:sec:NFW}). All physical quantities are expressed in these units. 
For the global profile $\Sigmabub$ in equation~\eqref{eq:dotnbub}, we adopt the following functional form
\begin{equation}\label{eq:SigmabubInR}
	\Sigmabub(r; \Rdisc) = \frac{1}{1 + r/ \Rdisc} \, .
\end{equation}
For ${r/\Rdisc \ll 1}$, it approximates the more realistic exponential disc, ${\Sigmabub \propto \re^{-r/\Rdisc}}$, and its simple functional form allows the orbit-averaging to be evaluated fully analytically.

Each \SB\@ is initially characterised by its maximum radius $\Rbub$, lifespan $\Tbub$, and mass density $\rhobub$ (see \appref{app:sec:ModellingSuperbubbles} for details on the modeling procedure). Alternatively, we hereafter mainly use the re-parameterized set of parameters ${(\Rbub, \Tbub, \Ebub)}$, where $\Ebub$ is the characteristic specific energy of a \SB\@.

Because our theoretical results are derived in the perturbative regime, they do not depend on any particular choice of \SB\@ parameters, provided these remain within that regime. For the comparison with numerical simulations in \secref{sec:NumericalValidation}, however, we adopt the ``fiducial'' set of values as ${\Rbub/\Rhalo = 0.25}$, ${\Tbub/\thalo = 0.25}$, ${\rhobub/\rhohalo = 0.1}$, ${\Rdisc /\Rhalo = 4.0}$, and ${\dotnbubz \, \Rhalo^{3} \thalo = 1.0}$ (see \appref{app:sec:Parameters} for their justifications). 

\begin{table}[!hbtp]	
	\centering
	\caption{Description of parameters adopted in this work.}
	\label{tab:Parameters}
	\begin{threeparttable}
		\small
		\setlength{\defaultaddspace}{0.25em}
		\begin{tabular*}{0.9\columnwidth}{@{\extracolsep{\fill}}ll}
			\toprule
			Galaxy parameter & Description \\
			\midrule
			$\Rhalo$ 	 	    & \DM\@ halo scale radius\tnote{a} \\ \addlinespace
			$\Vhalo$               & \DM\@ halo characteristic velocity\tnote{a} \\ \addlinespace
			$\thalo$     	    & \DM\@ halo timescale, ${\thalo \equiv \Rhalo/ \Vhalo}$  \\ \addlinespace
			$\rhohalo$            & \DM\@ halo mass density\tnote{a}  \\ \addlinespace
			$\Rdisc$               & Stellar disc scale radius\tnote{b} \\
			\midrule
			\SB\@ parameter & Description \\
			\midrule
			$\Rbub$ 	    	  & Maximum \SB\@ radius\tnote{c} \\ \addlinespace
			$\Tbub$ 	    	  & \SB\@ lifespan\tnote{c}   \\ \addlinespace
			$\rhobub$    	  &  \SB\@ mass density\tnote{c} \\ \addlinespace
			$\dotNbub$  	  & \begin{tabular}[t]{@{}l@{}} 
								Total \SB\@ formation rate within a sphere \\
						  		of radius $\Rdisc$,  ${\dotNbub = \dotnbubz \, \frac{4\pi}{3} \Rdisc^{3}}$\tnote{d}
						      \end{tabular} \\ \addlinespace
			$\dotNeff$    	&   \begin{tabular}[t]{@{}l@{}} 
								Effective \SB\@ formation rate,\\
								${\dotNeff \equiv A \dotNbub \left( \frac{\Rhalo}{\Rdisc}\right)^3 = A \dotnbubz \frac{4\pi}{3} \Rhalo^3}$\tnote{e}
						      \end{tabular}\\ \addlinespace			      
			$\Mbub$      	 & Characteristic \SB\@ mass, ${\Mbub \equiv \frac{4\pi}{3} \rhobub \Rbub^{3}}$ \\ \addlinespace
			$\Ebub$       	 &  \begin{tabular}[t]{@{}l@{}}
								Characteristic \SB\@ specific energy,  $\Ebub\!\equiv\! \frac{G \Mbub}{\Rbub} $
						     \end{tabular} \\ \addlinespace
			\midrule
			Diffusion quantity & Description \\
			\midrule
			$\tbal$				          & Ballistic timescale of perturbations\tnote{f}\\ \addlinespace
			$\tdyn$				          & Dynamical timescale of the \DM\@ orbit\tnote{f}\\ \addlinespace
			$\trel$				          & Relaxation timescale of the \DM\@ halo\tnote{f}\\ \addlinespace
			$\calE$ 		   		   	  & Shifted energy, ${\calE \equiv E + \Vhalo^{2}}$ \\ \addlinespace
			$D_{q_{i} q_{j}}^{\text{approx}}$  & \begin{tabular}[t]{@{}l@{}} 
										Approximate scaling relation for the diffusion \\
										 coefficients in ${q \in \{E, L\}}$ over all $E$ and $L$\tnote{g}
									     \end{tabular} \\ \addlinespace
			$D_{q_{i} q_{j}, 0}$  			  & Corresponding amplitude of $D_{q_{i} q_{j}}^{\text{approx}}$\tnote{g} \\ \addlinespace					
			$\calEmax$ 			   & \begin{tabular}[t]{@{}l@{}} 
										Corresponding energy at maximum \\
										diffusion of $D_{q_{i} q_{j}}^{\text{approx}}$\tnote{g}
									      \end{tabular} \\				      
			\bottomrule
		\end{tabular*}
		\begin{tablenotes}
			\scriptsize
			\item[a] See equations~\eqref{eq:NFWPotential}--\eqref{eq:NFWDensity}. 
			\, \tnote{b} See equation~\eqref{eq:SigmabubInR}.
			\, \tnote{c} See \appref{app:sec:ModellingSuperbubbles}.
			\item[d]  Here, $\dotnbubz$ denotes the central creation rate per unit volume in equation~\eqref{eq:dotnbub}. 
			\item[e] The numerical constant $A$ follows from the choice of power spectra (see \appref{app:sec:ScalingRelation}).
			\, \tnote{f} See \secref{sec:Theory}.
			\, \tnote{g} Derived in this work; see \secref{subsec:ScalingRelations}.
		\end{tablenotes}
	\end{threeparttable}
\end{table}
\subsection{Radial orbits}\label{subsec:RadialOrbits}
\begin{figure}[!htbp]
	\center \includegraphics[width=0.48\textwidth]{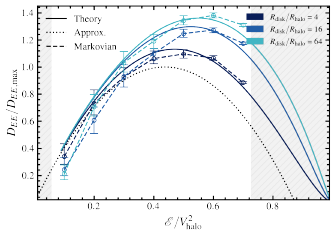}
	\caption{ 
	 $D_{EE}$ from the theoretical prediction (solid lines), the approximate scaling relation from equation~\eqref{eq:DEEofEScaling} (black dotted line), and the \Mkv\@ simulation (dashed lines with data points), all scaled by $\DEEmax$, given by equation~\eqref{eq:DEEMax}. Different $\Rdisc$ parameters are coded by colors in the legend. The gray hashed region represents the limit beyond the size of the simulation box (right), and below the grid size (left). For ${\Rdisc / \Rhalo= 4}$, the three results overlap and show better agreement in the tightly bound limit ${\calE \to 0}$. But for ${\Rdisc/\Rhalo = 16}$ and 64, increasing the disc scale radius $\Rdisc$ shifts both $\calEmax$ and $\DEEmax$, indicating that a more extended distribution of \SBs\@ leads to stronger diffusion in the less-bound regime.
	}
	 \label{fig:DEEofE}
\end{figure}
In this section, we first consider radial orbits for two reasons, although the velocity distribution of the inner \DM\@ halo is close to being isotropic \citep{HansenMoore2006}. First, radial orbits are the limiting case of highly eccentric orbits, which repeatedly pass through the central cusp and are therefore most sensitive to the potential fluctuations concentrated at the centre \citep{Pontzen2012,ElZant2016,BurgerZavala2021}. Moreover, their strongly anharmonic radial motion spreads over many orbital harmonics, probing a broad range of fluctuation frequencies (see \figref{fig:PowerSpectrum}; \citealt{Ogiya2014}). Second, in the perturbative regime, this limit allows the diffusion coefficients to be computed fully analytically, 
yielding approximate scaling relations that extend to more general orbits in \secref{subsec:GeneralOrbits} and in \secref{subsec:ScalingRelations}. 
 
For radial orbits, \figref{fig:DEEofE} shows the diffusion coefficient in energy, $D_{EE}$, as a function of the shifted energy, defined as $\calE \equiv E + \Vhalo^{2}$. The diffusion coefficient is scaled by its maximum value $\DEEmax$, given by equation~\eqref{eq:DEEMax} below. From the theoretical prediction (solid lines), $D_{EE}$ exhibits a bell-shaped profile, where the diffusion in energy vanishes at the boundaries. In the  ${\calE/\Vhalo^{2} \to 0}$ limit, the orbit is confined in the centre of the \DM\@ halo, where potential fluctuations by \SBs\@ are significantly outweighed by the mean-field potential. Conversely, the number of perturbers experienced by the orbit becomes negligible in the limit ${\calE/\Vhalo^{2} \to 1}$. As a result, the diffusion in energy is maximized at intermediate $\calE$, in which the \DM\@ orbit both encounters a sufficient number of \SBs\@ and is not strongly suppressed by the strength of the mean-field \DM\@ potential. A shift in $\DEEmax$ and $\calEmax$ is also observed when varying the inhomogeneity of \SBs\@. Increasing $\Rdisc$ (color coded) distributes more \SBs\@ towards larger radii, enhancing the overall diffusion. Along the same lines, at a given energy $\calE$, the orbit effectively encounters more \SBs\@ in a more extended disc, leading to an increase in $\calEmax$.

We emphasize that the normalized diffusion coefficients in energy, ${D_{EE}/\DEEmax}$, predicted by the full theoretical model (solid lines in \figref{fig:DEEofE}) and \Mkv\@ simulation (dashed lines in \figref{fig:DEEofE}; see \secref{sec:NumericalValidation}) closely overlap. This agreement indicates that the theory successfully captures both the energy dependence on its shape and its amplitude in terms of characteristic parameters. This establishes a  foundation for deriving the proper scaling of diffusion, thereby enabling quantitative predictions of its dependence on the relevant physical parameters (see \secref{subsec:ScalingRelations}).

\subsection{General orbits}\label{subsec:GeneralOrbits}
\begin{figure}[!htbp]
\begin{subfigure}[b]{0.50\textwidth}
	\centering
	\caption*{(a) $D_{EE}$}
	\includegraphics[width=1.0\textwidth]{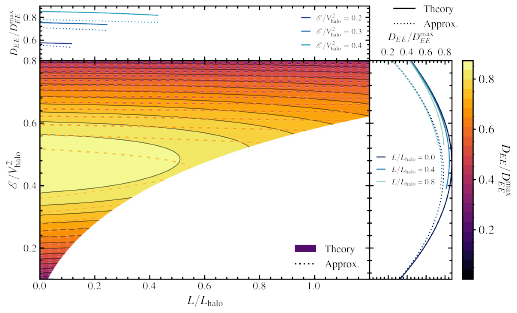}
\end{subfigure}
\begin{subfigure}[b]{0.50\textwidth}
	\centering
	\caption*{(b) $D_{LL}$}
	\includegraphics[width=1.0\textwidth]{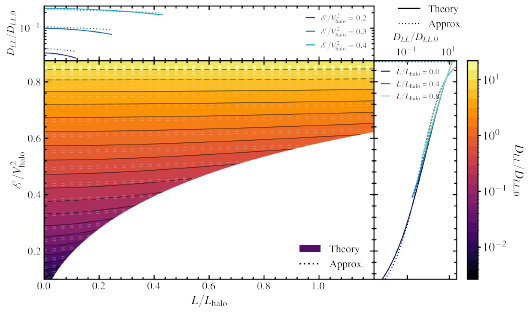}
\end{subfigure}
\begin{subfigure}[b]{0.50\textwidth}
	\centering
	\caption*{(c) $D_{EL}$}
	\includegraphics[width=1.0\textwidth]{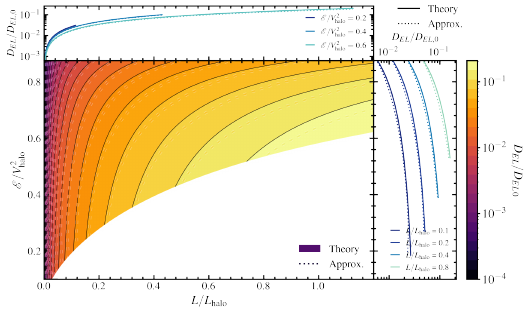}
\end{subfigure}
	\caption{Theoretical predictions of the orbit-averaged diffusion coefficients $D_{EE}$, $D_{LL}$, and $D_{EL}$ (top to bottom panels), scaled by the normalization factors defined in equation~\eqref{eq:D0sets}, for general orbits embedded in a spherically symmetric distribution following equation~\eqref{eq:SigmabubInR} with $\Rdisc/\Rhalo = 4$. The normalized ratios $D_{EE} / D_{EE}^{\text{max}}$, $D_{LL} / D_{LL,0}$, and $D_{EL} / D_{EL,0}$ are indicated by the color maps. Line contours depict the approximate scalings given by equation~\eqref{eq:ScalingRelation}. In each panel, the top (resp. right) marginal panel displays the corresponding diffusion coefficient as a function of angular momentum $L$ at fixed energy $\calE$ (resp. as a function of energy $\calE$ at fixed angular momentum). Overall, the approximate scaling relations capture both the absolute magnitude of the diffusion coefficients and their variation across $(\calE, L)$ phase space, providing an intuitive description of the diffusion process.
	 }
	\label{fig:DTheory3D}
\end{figure}
\figref{fig:DTheory3D} presents one of the main results of this work: the theoretical predictions for $D_{EE}$, $D_{LL}$, and $D_{EL}$ as functions of ${(\calE, L)}$ in a spherically symmetric halo. For energy diffusion (top panel), the normalized coefficient, $D_{EE} / \DEEmax$, retains a bell-shaped dependence on $\calE$, while its amplitude decreases only modestly with increasing angular momentum. This indicates that, at a fixed energy, energy diffusion depends only weakly on $L$; consequently, orbital anisotropy exerts a minor effect on energy transfer (see also \secref{subsec:DensityEvolution}). Furthermore, the peak energy $\calEmax$ of $D_{EE}$ is virtually independent of $L$.

For diffusion in angular momentum (middle panel), the normalized diffusion coefficient, $D_{LL} / D_{LL, 0}$, increases approximately quadratically with energy. This trend reflects the dependence of angular momentum changes on perturbations in position and velocity, $(\br, \bv)$, given that $(\delta \bL)^{2} \propto [\delta (\br \times \bv)]^{2}$. Conversely, the overall functional shape of the normalized coefficient varies only weakly with $L$. However, at a fixed energy, a given absolute change in angular momentum corresponds to a larger fractional change, $\delta L / L$, for orbits with smaller initial $L$. Thus, orbits with small initial $L$ undergo more substantial relative changes, even though the absolute diffusion coefficient varies weakly with $L$. Although eccentric orbits reach larger apocentric radii, where perturbing torques are stronger than near the centre, the resulting effective angular momentum diffusion depends critically on the global spatial distribution of the perturbations. Because we adopt a mildly declining profile such as equation~\eqref{eq:SigmabubInR}, the angular momentum scaling in ${D_{LL}^{\mathrm{approx}}}$ is weaker than it would be for a steeper distribution, such as an exponential profile.

Finally, the cross-diffusion coefficient in $(E, L)$ characterises the covariance between changes in orbital energy and angular momentum. The normalized coefficient, $D_{EL} / D_{EL, 0}$, vanishes in the radial orbit limit ($L \to 0$), because deflections in $E$ and $L$ decouple along purely radial trajectories. As angular momentum increases at fixed energy, approaching the circular orbit limit, the cross-term grows more pronounced, indicating a stronger correlation between energy and angular momentum variations. Its contribution to the evolution of the \DF\@ is governed by the mixed derivative term in the \FP\@ equation. In particular, ${\p^{2} D_{EL} / \p E \p L}$ changes sign around $\calE \simeq 4\calEmax / 3$ in our model, meaning the cross-term contributes with opposite signs on either side of this transitional energy.

\subsection{Scaling relations}\label{subsec:ScalingRelations}
In the limit of ${\calE/\Vhalo^{2} \ll 1}$, ${\Rbub / \Rhalo \ll 1}$, ${\Tbub/\thalo \ll 1}$, and ${\Rdisc/\Rhalo \gg 1}$, a scaling relation holds for the amplitude of the diffusion coefficients, $\Dz$;  this motivates the corresponding dimensional normalizations for the second-order diffusion coefficients
\begin{align}\label{eq:D0sets}
	\begin{bmatrix}
	D_{EE,0} \\ \addlinespace
	D_{LL,0} \\ \addlinespace
	D_{EL,0}
	\end{bmatrix}
	&= 
	\overbrace{
	\dotNeff
	\left( \frac{\Ebub}{\Vhalo^{2}} \right)^{2}
	\left(\frac{\Rbub}{\Rhalo}\right)
	\left( \frac{\Tbub }{\thalo} \right)
	}^{\displaystyle \Dz \equiv }
	\begin{bmatrix}
	\Vhalo^{4} \\ \addlinespace
	5\Rhalo^{2} \Vhalo^{2} \\ \addlinespace
	4\Rhalo \Vhalo^{3}
	\end{bmatrix} \, .
\end{align}
By further introducing the angular momentum of the circular orbit at energy $E$ and circular radius $\rc$ as
\begin{equation}\label{eq:Lc}
	\Lc  = \rc\,  \sqrt{2[E - \Phi(\rc)]} \, ,
\end{equation}
and the energy at which the diffusion in energy approximately reaches its maximum,
\begin{equation}\label{eq:EMaxApproxPade}
	\calEmax = \frac{\Vhalo^{2}}{2}\left[1- \frac{1}{2} \left(\frac{\Rhalo}{\Rdisc}\right) \right] \, ,
\end{equation}
we obtain the following approximate scaling relations for the second-order diffusion coefficients
\begin{subequations}\label{eq:ScalingRelation}
\begin{align}
D_{E E}^{\text{approx}} &= D_{E E, 0} \, \left( \frac{\calE}{\Vhalo^{2}} \right) \left( 1 - \frac{\calE}{2 \calEmax} \right) \left(1- \frac{L}{8\Lc[\calE]} \right)^{1/4} \, , \label{DEEApprox}\\
D_{L L}^{\text{approx}} &= D_{L L, 0} \, \left( \frac{\calE}{\Vhalo^{2}} \right)^{\!2}  \left( 1 - \frac{\calE}{2 \calEmax} \right)^{\!-1}  \left(1- \frac{L}{4\Lc[\calE]} \right)^{1/2} \, , \\
D_{EL}^{\text{approx}} &= D_{E L, 0} \, \left( \frac{\calE}{\Vhalo^{2}} \right)^{\!2} \left( 1 - \frac{\calE}{2 \calEmax} \right) \left(\frac{L}{\Lc[\calE]} \right) \, .
\end{align}
\end{subequations}
These approximate expressions are represented in dotted lines in \figref{fig:DEEofE} and \figref{fig:DTheory3D}.

We highlight that equations~\eqref{eq:ScalingRelation} encapsulate a quantitative description of the functional forms of the dependence on perturber's characteristic parameters ($\Rbub,\Tbub, \Ebub$ and $\dotNeff$ via $D_0$, and $ \Rdisc$) and orbital labels ($\calE$ and $L$). This is one of the main results of this work. We refer to \appref{app:sec:ScalingRelation} for more details on how to derive the scaling and how well it holds.

In the radial orbit limit ${L \to 0}$, equation~\eqref{DEEApprox} reduces to
\begin{equation}\label{eq:DEEofEScaling}
	D_{EE}^{\text{approx, rad}} (\calE) = 
	D_{EE, 0}\, 
	\left(
		\frac{\calE}{\Vhalo^{2}}
	\right)
	\left(
		1 - \frac{\calE}{2\calEmax}
	\right),
\end{equation}
and the maximum diffusion rate is given by plugging equation~\eqref{eq:EMaxApproxPade} into equation~\eqref{eq:DEEofEScaling} as
\begin{equation}\label{eq:DEEMax}
	\DEEmax =  \frac{D_{EE, 0}}{4} \left[1- \frac{1}{2} \left(\frac{\Rhalo}{\Rdisc}\right) \right]\,,
\end{equation}
where $D_{EE, 0}$ was defined in equation~\eqref{eq:D0sets}.
Equation~\eqref{eq:DEEofEScaling} is shown as a dotted line on Figure~\ref{fig:DEEofE}.

From equations~\eqref{eq:D0sets} and~\eqref{eq:DEEMax}, one can infer that the amplitude of the maximum energy diffusion scales with the squared energy ratio between the energy injected by \SBs\@ and the mean-field energy of the halo. This ratio reflects the required commensurability between the halo and \SB\@ parameters for the coupling to be efficient. The associated scaling depends on both the geometry of \SB\@, which determines its fluctuation power spectrum, and also on the net impact of orbit-averaging. Similarly, diffusion is maximized when the mean-field potential and the perturber's features \emph{resonate} at the relevant scales: when the characteristic size $\Rbub$ is commensurate with $\Rhalo$, and the timescale $\Tbub$ to $\thalo$.

The scaling relations given by equation~\eqref{eq:ScalingRelation} faithfully reproduce the structure of the diffusion coefficients, as illustrated by the contour lines in \figref{fig:DTheory3D}. As a key result of this work, these analytic expressions can be readily reused in other diffusion processes in astrophysics, and we will present further applications in a companion paper. Retaining the normalization factors defined in equation~\eqref{eq:D0sets}, the parabolic form in $(\calE, L)$ only captures the theoretical predictions across phase space. In particular, the energy diffusion coefficient shows good agreement between the exact theory and the analytic approximation in the low-energy regime, ${0.1 \lesssim \calE / \Vhalo^{2} \lesssim 0.3}$ (see the right marginal panel of the top row in \figref{fig:DTheory3D}). In this regime, the orbits reside sufficiently deep within the halo potential (see \appref{app:sec:ScalingRelation}), while remaining far enough from the centre for the local treatment of perturbations to remain valid.

\section{Numerical validation}\label{sec:NumericalValidation}
\begin{figure}[htbp]
\begin{subfigure}[b]{0.50\textwidth}
	\centering
	\caption*{(a) $D_{EE}$}
	\includegraphics[width=1.0\textwidth]{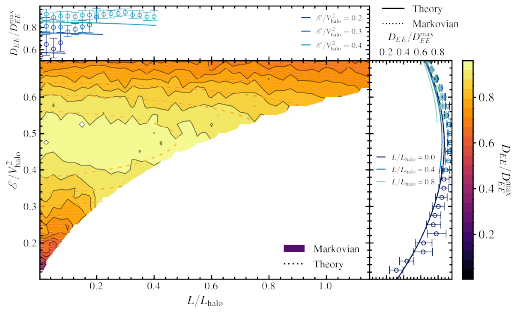}
\end{subfigure}
\begin{subfigure}[b]{0.50\textwidth}
	\centering
	\caption*{(b) $D_{LL}$}
	\includegraphics[width=1.0\textwidth]{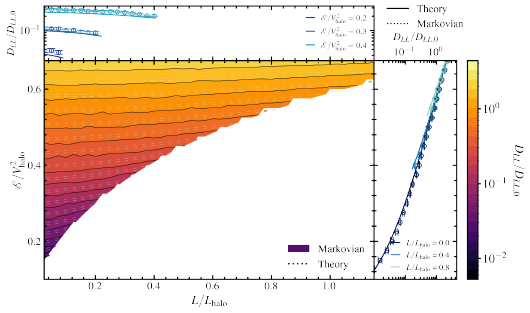}
\end{subfigure}
\begin{subfigure}[b]{0.50\textwidth}
	\centering
	\caption*{(c) $D_{EL}$}
	\includegraphics[width=1.0\textwidth]{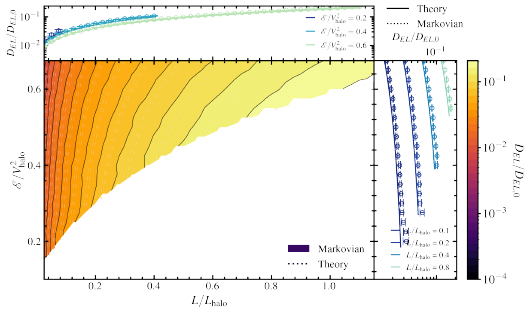}
\end{subfigure}
	\caption{
	Measured diffusion coefficients $D_{EE}$, $D_{LL}$, and $D_{EL}$ (from top to bottom panel) in \Mkv\@ simulations for the \DM\@ halo and \SB\@ parameters equal to those of \figref{fig:DTheory3D}. Line contours are overlaid from the theoretical prediction. Across all $(\calE, L)$ bins, the measurements agree with the theoretical prediction within 1--5$\%$.
	}
	\label{fig:DMarkovian3D}
\end{figure}
Let us now validate the theoretical orbit-averaged diffusion coefficients against two distinct sets of numerical simulations: (i)~\Mkv\@, and (ii)~\nMkv\@ simulations. To summarise, we generate $128$ realizations of background \SBs\@ and $1\,024$ initial conditions of \DM\@ orbits, at a given ${(E, L)}$, and evolve the \DM\@ orbital trajectory for those initial conditions. We then track the deviations ${(\delta E[t], \delta L[t])}$ as a function of time $t$. 

The overall procedure is the same for both simulations, but the driving force governing the orbital evolution partially differs. In the \Mkv\@ case, the local deviations are sampled while the trajectory is determined only by the mean-field potential $\Phi$, neglecting the potential fluctuations by \SBs\@, $\delta \Phi$. In contrast, in the \nMkv\@ case, the orbital evolution is sourced by the combined potential ${\Phi + \delta \Phi}$. We focus here on the main result, while deferring the detailed methodology to \appref{app:sec:NumericalSimulation}.

Overall, the diffusion coefficients measured in the \Mkv\@ simulation agree with our theoretical predictions (overlaid contours) to within 1--5$\%$ of the uncertainties, as shown in $(E, L)$ space for ${\Rdisc / \Rhalo = 4}$ in \figref{fig:DMarkovian3D}. This close agreement is expected, as the \Mkv\@ setup directly satisfies our theoretical assumptions. However, a noticeable divergence emerges as ${\calE/\Vhalo^{2} \to 0}$, where steeper potential gradients near the halo centre invalidate the local perturbation approximation (see \secref{subsec:Limitations} for further discussion).

For the \nMkv\@ simulation, the measured diffusion coefficients reproduce the overall shape of the theoretical prediction, but their amplitude is larger by a factor of a few. This boost decreases systematically as $\Rbub$ and $\Tbub$ are reduced: the ratio of the \nMkv\@ to the \Mkv\@ coefficients converges to unity, while the shape remains intact. This may indicate that the theory based on two-body relaxation captures fairly well the phase-space structure of the diffusion, even when the \Mkv\@ assumption is relaxed,  and that \nMkv\@ effects mainly rescale its amplitude. 
Interestingly, \cite{Tep2022} reached a similar conclusion when comparing the orbit-averaged Chandrasekhar prediction with the inhomogeneous Landau prediction for the self-induced relaxation of globular clusters.
We detail the results of \nMkv\@ simulations and their convergence to the \Mkv\@ simulation in \appref{app:sec:ValidityOfMarkovianAssumption}, and discuss the plausible origin of the amplitude mismatch in \secref{subsec:Limitations}.

\section{Cusp-to-core transformation}\label{sec:CuspCoreTransformation}
Once the diffusion coefficients (e.g., equation~\ref{eq:ScalingRelation}) and the initial \DF\@ are specified, the mean evolution of physical quantities of interest (such as the density profile and the characteristic heating timescale of the \DM\@ halo) can be obtained by integrating the \FP\@ equation~\eqref{eq:FokkerPlanckInEL}. In \secref{subsec:Galaxies}, we estimate the characteristic timescale over which dynamical heating occurs and express it in physical units using parameters directly constrained by observations. This provides a quantitative metric for the rate of secular diffusion as a function of the perturbers' parameters. In \secref{subsec:DensityEvolution}, we present the temporal scaling of the inner density slope of an \NFW\@ profile and validate it by numerically integrating the \FP\@ equation. We refer the reader to \appref{app:sec:NFW} for details on the \DF\@ models of \NFW\@ haloes, and to \appref{app:sec:EvolutionOfDensityProfile} for the numerical methods used to solve the \FP\@ equation with approximate diffusion coefficients.
\subsection{Diffusion timescale of MW, dwarf and giant galaxies}\label{subsec:Galaxies}
Let us define a diffusion timescale, $\tdiff$, at maximum diffusion in energy (e.g., \S{7.4.5} in \citetalias{BinneyTremaine2008}) in units of the halo dynamical timescale $\thalo$ as
\begin{align}\label{eq:tD}
\frac{\thalo}{\tdiff} &\equiv \frac{\DEEmax}{\Vhalo^{4}} \thalo \\
&=  \frac{1}{4} ( \dotNeff \thalo) \,
	\left( \frac{\Ebub}{\Vhalo^{2}} \right)^{2}
	\left(\frac{\Rbub}{\Rhalo}\right)
	\left( \frac{\Tbub }{\thalo} \right)
	 \left[1- \frac{1}{2} \left(\frac{\Rhalo}{\Rdisc}\right) \right].  \notag
\end{align}
From equation~\eqref{eq:tD}, we can compute the maximum diffusion timescales in physical units for dwarf and \MW\@-like galaxies
\begin{align}\label{eq:tDwithNumbers}
	\tdiff &\approx 
	\left( \frac{1.1 \times 10^{45} \, \erg \, \Msol^{-1}}{\Ebub} \right)^{2}
	\left( \frac{250 \, \pc}{\Rbub} \right)
	\left( \frac{25 \, \Myr}{\Tbub } \right) \notag \\
	&\qquad\qquad{}
	\times \begin{dcases}
	2.0 \left( \frac{10^{4}  \, \Msol \, \Myr^{-1}}{\overline{\SFR}} \right) \, \Gyr, & \text{for dwarf},  \\
	10.5 \left( \frac{2 \times 10^{6} \, \Msol \, \Myr^{-1}}{\overline{\SFR}} \right) \, \Gyr, & \text{for \MW\@},
	\end{dcases}
\end{align}
where $\overline{\SFR}$ is the mean \SFR\@ for a halo of the corresponding mass, from which $\dotNbub$ is derived, and the adopted numbers are motivated in \appref{app:sec:Parameters}. Note that this estimate should be considered as an upper limit on the diffusion timescales, although we argue that it remains within a reasonable range; see \secref{subsec:Limitations} for a detailed justification.

For dwarf galaxies, the resulting diffusion timescale of ${2.0 \, \Gyr}$ in equation~\eqref{eq:tDwithNumbers} is comparable to the timescales over which cusp-to-core transformations operate as found in previous studies \citep{Teyssier2013,Read2016,Muni2024,Hayashi2025}. Repeated episodes of energetic stellar feedback in this halo mass range have indeed been shown to transform an initially cuspy \DM\@ profile (${\gamma \gtrsim 1}$; see equation~\ref{eq:GeneralizedNFWDensity}) into a substantially shallower, cored one (${\gamma \sim 0.4}$) over timescales of approximately ${2.0}$--${3.5 \,\Gyr}$ in previous observations and simulations \citep{Oh2011,DiCintio2014}, and analytic predictions \citep{Pontzen2012}. 
 
As expected for more massive \MW\@-like galaxies, the predicted diffusion timescale is considerably longer, reaching approximately ${10.5 \, \Gyr}$ for the adopted parameters. This longer timescale suggests that diffusion is less efficient at modifying the inner density profile over a given period, maintaining the cuspy profile against ongoing perturbations compared to dwarf galaxies. This trend is also broadly consistent with the mass dependency of feedback-driven cusp-to-core transformations found in simulations \citep{Jackson2024}. 

At moderate stellar-to-halo mass ratios and star formation thresholds, the efficiency of energy transfer to \DM\@ via recurrent feedback-driven outflows increases, leading to the formation of central cores. However, above a critical threshold in the central baryonic mass, this heating competes with the deepening central gravitational potential, causing the halo to undergo renewed central contraction and re-establish a cusp \citep{DiCintio2014,Tollet2016}. In line with our prediction in equation~\eqref{eq:tDwithNumbers}, although the elevated star formation parameters encoded in $\dotNbub$ increase the rate of perturbations, their cumulative impact must compete against an increasingly dominant central potential. Thus, the longer diffusion timescales in more massive galaxies offer a natural explanation for the persistence of cuspy halo profiles at higher galaxy masses.

For giant/cD galaxies, we instead consider bubbles driven by \AGN\@ feedback, replacing the \SB\@ parameters with their corresponding \AGN\@ counterparts. This substitution is motivated by the fact that both stellar and \AGN\@ feedback involve episodic mass outflows, although their underlying microphysics and characteristic physical scales differ (see \appref{app:subsec:Superbubble} for further justification). Assuming that an \AGN\@ bubble has a larger radius of 
${R_{\text{AGN}} \sim 30\, \kpc}$, longer lifetime ${T_{\text{AGN}} \sim 100\, \Myr}$, energy scale ${E_{\text{AGN}} \sim 3 \times10^{59} \, \erg }$, recurrence rate ${\dot{N}_{\text{AGN}} \sim T_{\text{AGN}}^{-1} = 10 \, \Gyr^{-1}}$, and is concentrated near the galactic centre ${\Rdisc/\Rhalo \sim 1}$, the diffusion timescale is given by
\begin{align}\label{eq:tDwithAGN}
	\!\! \tdiff &\!\approx\! 
	6.5
	\left( \frac{3 \times 10^{59} \, \erg }{E_{\text{AGN}}} \right)^{2}\!
	\!\left( \frac{30 \, \kpc}{R_{\text{AGN}}} \right)
	\!\left( \frac{100 \, \Myr}{T_{\text{AGN}} } \right)
	\!\left( \frac{10 \, \Gyr^{-1}}{\dot{N}_{\text{AGN}}} \right) \, \Gyr .
\end{align}

\subsection{Evolution of inner density slope $\gamma$}\label{subsec:DensityEvolution}
\begin{figure*}[!htbp]
	\centering
\begin{subfigure}[t]{0.45\textwidth}
		\centering
		\includegraphics[width=0.90\textwidth]{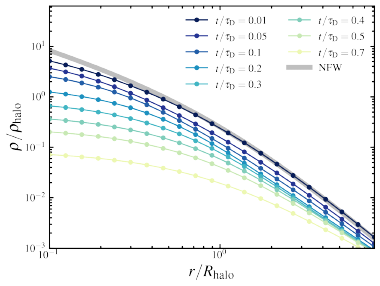}
		\caption{Evolution of the \NFW\@ density profile}
		\label{fig:DensityEvolution}
	\end{subfigure}
\begin{subfigure}[t]{0.44\textwidth}
		\centering
		\includegraphics[width=0.90\textwidth]{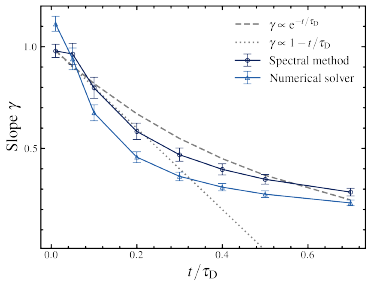}
		\caption{Evolution of the measured and predicted inner density slope}
		\label{fig:GammaEvolution}
	\end{subfigure}
\caption{Left panel (\subref*{fig:DensityEvolution}): evolution of the \NFW\@ density profile at different time steps (color coded) for an isotropic halo system. The thick gray line represents the original \NFW\@ profile, given by equation~\eqref{eq:NFWDensity} with inner density slope ${\gamma = 1}$. The cusp-to-core transformation occurs over a diffusion time, as expected. Right panel (\subref*{fig:GammaEvolution}): evolution of the measured and predicted inner density slope, $\gamma$, as a function of time (scaled by the diffusion timescale). The dashed line depicts the theoretical prediction based on the dominant eigenvalue, whereas the dotted line shows its early-time linear approximation. Open circles (resp.\ open triangles) denote the measured inner density slope obtained from a linear regression of ${\log \rho}$ against ${\log r}$ over ${r / \Rhalo \le 0.5}$ using the spectral  (resp.\ direct numerical integration) solution. Error bars represent the standard deviations estimated from the covariance matrix of the linear fits.	
	The expected cusp-to-core transformation is reproduced by the model.
	\label{fig:FPEvolution}
	 }
\end{figure*}
By assuming a separable-in-time solution to the \FP\@ equation, equation~\eqref{eq:FokkerPlanckInEL}, we can deduce a simple scaling for the evolution of the inner density slope via a spectral decomposition. Assuming that the early-time evolution ($t \ll \tdiff$) is governed by the slowest-decaying mode (corresponding to the largest non-zero and strictly negative eigenvalue), we obtain the following scaling for the inner density slope (see \appref{app:sec:EvolutionOfDensityProfile})
\begin{equation}\label{eq:GammaScalingIsotropicDF}
\frac{\dot{\gamma}}{\gamma} \propto -\frac{1}{\tdiff}.
\end{equation}
This indicates $\gamma(t) \propto \re^{-t/\tdiff}$ so that $\gamma$ decreases linearly in time at early stages.
We illustrate the change in the inner region of the density profile for an isotropic halo system and the corresponding decrease in inner density slope in Figure~\ref{fig:FPEvolution}.

To summarize our main results, we have developed a framework describing the secular evolution of a dominant mean-field system (\DM\@ halo) subject to recurrent, stochastic baryonic perturbations (\SBs\@), valid in the perturbative regime. Specifically, this work (i)~incorporates realistic astrophysical power spectra (see \figref{fig:PowerSpectrum}), (ii)~accounts for global spatial inhomogeneity, moving beyond the standard homogeneous approximation (see \secref{subsec:GlobalPhaseAveraging}), and (iii)~yields analytic and physically transparent results (see equations~\ref{eq:D0sets} and \ref{eq:ScalingRelation}). This formalism is sufficiently versatile to model a broad range of perturbing sources, from perturbers of arbitrary spatial and temporal extents to diverse forms of baryonic feedback, while providing a quantitative framework to predict how their cumulative effects scale over time.

\section{Discussion}\label{sec:Discussion}
\subsection{Prior works}
Several past studies have considered idealized prescriptions of the time-dependent potential fluctuations associated with baryonic feedback. For example, \citet{Pontzen2012} introduced a time-varying harmonic oscillator model in which rapid, recurrent potential fluctuations driven by supernova-driven gas blowout transfer energy to the \DM\@ component. Alternatively, \citet{Ogiya2014} modelled periodic potential variations and emphasized the role of orbital resonances between \DM\@ particles and perturbation waves, showing that energy transfer is maximized when the corresponding timescales coincide. While these idealizations yield valuable insights into cusp-to-core transformations, perturbations generated by realistic baryonic feedback are inherently more complex than a single harmonic or periodic mode. A realistic treatment must therefore account for the broad spectrum of amplitudes and frequencies characteristic of real feedback-driven perturbers. Furthermore, over secular timescales, a \DM\@ orbit samples fluctuations across diverse spatial locations and orbital phases, introducing stochastic phase modulations to its evolution. This motivates the ensemble-averaged kinetic framework developed in this paper. It systematically captures both the cumulative impact of repeated deflections and the contributions from a continuum of perturbation frequencies.

A statistical approach to this problem was considered by \citet{ElZant2016}, who investigated cusp-to-core transformations by coupling \DM\@ orbits to homogeneous gas fluctuations modelled as Gaussian random fields. Although their model captures stochastic fluctuations, its scale-free assumption omits spatial inhomogeneities as well as the specific spectral signatures characteristic of feedback-driven density perturbations. In realistic galaxies, both the spatial distribution and power spectrum of these fluctuations deviate substantially from such an idealized picture. In this context, the orbit-averaging framework offers a complementary method to bridge local, scale-dependent power spectra with the cumulative secular impact of realistic feedback-driven perturbations across the galaxy.

Recently, \citet{Kipper2025} used time-dependent \SB\@ structures motivated by cosmological simulations and observations to investigate their impact on dynamical friction. They constructed a locally evolving density profile grounded in galaxy formation models to calculate how gas processes couple to \DM\@ orbits through linear and angular momentum transfer. Their formulation assumed a globally homogeneous distribution of \SBs\@, adhering to the classical framework of \citet{Chandrasekhar1943}, and focused primarily on the deterministic acceleration field. Our work complements that of \citet{Kipper2025} by determining not only the mean drift of the orbital labels $(E, L)$, but also their full stochastic diffusion across phase space.

 Using magnetohydrodynamic simulations from the TIGRESS-NCR framework, \citet{Modak2026Characterizing} measured the power spectrum of the interstellar medium, capturing a qualitative behaviour similar to the bell-shaped profile in our model (\figref{fig:PowerSpectrum}), though lacking the turnover at large spatial scales (see Figure~8 and the discussion in \S{5.2} therein). Indeed, they note that larger simulation domains are required to fully identify this spectral break.
  
 \citet{Weinberg2025} investigated the dissolution of the bar by baryonic feedback using the Hamiltonian mean-field model. In this context, they concluded that the large-scale central perturbation plays a more important role than the small-scale bubble-like structures.
Our approach instead focuses on the stochastic diffusion of \DM\@ orbits, which incorporates ensemble-averaged perturbations as well as orbit averaging.

\subsection{Limitations}\label{subsec:Limitations}
Although the \Mkv\@ simulation is designed to reproduce theoretical predictions by evolving orbital trajectories within the background mean-field potential, it does not always yield perfect agreement with analytical theory. In practice, finite numerical resolution limits the ability to sample the full power spectrum, inevitably introducing effective cut-offs. We further discuss the impact of frequency truncation and its implications in \appref{app:secTruncationInFrequency}.

Furthermore, evaluating the local diffusion coefficients relies on the assumption of an infinitesimal local domain, which stands in inherent tension with the aforementioned numerical resolution constraints. Although the \Mkv\@ simulation results agree well with theoretical predictions over a broad parameter space, noticeable deviations emerge when \SB\@ parameters differ substantially from their fiducial values, or in the weakly bound limit (${\calE / \Vhalo^{2} \to 1}$; see, e.g., \figref{fig:DEEofE} and \secref{sec:NumericalValidation}).

Note that these resolution-driven artifacts likely impact cosmological simulations with intermediate to low resolution, in which the secular rate of physically driven orbital diffusion can be underestimated \citep{FaroukiSalpeter1982,Diemand2004,HolleyBockelmann2005,VanDenBoschOgiya2018}. Analytical models such as those presented in  \appref{app:secTruncationInFrequency} can be used to quantify this effect.

Our theoretical phase averaging framework relies on the perturbative approach of weak and local deflectors. This requires the characteristic \SB\@ parameters to be small compared to those of the halo: ${\Rbub/ \Rhalo \ll 1}$, ${\Tbub/\thalo \ll 1}$, ${\rhobub / \rhohalo \ll 1}$. However, the validity of this assumption may break down in two regimes: (i)~when perturbations are non-local in space or time, and (ii)~when their amplitude becomes comparable to the mean-field potential. In such cases, the perturbative expansion and the diffusion description as a result of a \Mkv\@ process may no longer hold.

Despite this discrepancy, in that regime, we find remarkably that the \nMkv\@ results preserve the shape of diffusion, with an overall amplification in amplitude (e.g., \appref{app:sec:ValidityOfMarkovianAssumption}). This amplification factor decreases as the perturbations become more localized: a decrease in $\Rbub$, $\Tbub$, and $\rhobub$ (see \appref{app:sec:ValidityOfMarkovianAssumption} for details). One possible explanation for the remaining mismatch in amplitude is the absence of global resonant effects in our treatment \citep[e.g.][]{PichonAubert2006}. Incorporating such effects through a resonant relaxation formalism could cure the discrepancy and provide a more complete description of the long-term couplings between orbits and perturbers. We should also consider self-consistently the collective response of the system as it evolves. However, such approaches would likely compromise the analytic tractability of our framework, particularly the derivation of simple scaling relations such as equation~\eqref{eq:ScalingRelation}. Moreover, dynamically hot systems are mostly dominated by small-scale fluctuations, for which non-resonant descriptions have proven to be accurate
\citep[e.g.,][]{Theuns1996,Fouvry2021,Tep2026}.

Finally, we presented the diffusion coefficients for \DM\@ particles whose orbits lie within the stellar disc. Because this coplanar configuration generally minimizes the relative velocity between \DM\@ particles and \SBs\@, it maximizes the strength of individual encounters and the resulting diffusion \citep{Quinn1986,Weinberg1987,Penarrubia2002}. Thus, our main result provides an upper-limit estimate of the diffusion for orbits in a spherical halo. Nonetheless, we expect this approximation to remain reasonably representative of the inner halo for two reasons. First, inner halo orbits are more likely to intersect the stellar disc: given their tighter radial extent and shorter orbital periods, they cross the disc plane more frequently
with better orbital frequency match. Moreover, within the inner halo  (e.g., $\lesssim 0.1$--$0.2$ virial radius), the \DM\@ halo responds to and co-evolves with the baryonic disc, becoming more oblate and aligning with the disc spin axis \citep{Bailin2005,Abadi2010,Chua2019}. Notwithstanding, explicitly accounting for the three-dimensional axisymmetric configuration of the stellar disc alters the overall amplitude of the diffusion coefficients by only a factor of a few, while moderately modifying their distribution across 
$(\calE, L)$ (see \figref{fig:DTheoryAxisymmetric}). Therefore, although the coplanar approximation overestimates the diffusion experienced by orbits with significant out-of-plane motion, it still provides a useful benchmark for assessing the overall efficiency of feedback-driven diffusion.

\section{Conclusions and perspectives}\label{sec:ConclusionsAndPerspectives}

\subsection{Conclusion}\label{subsec:Conclusion}
Our main findings are the following:
\begin{itemize}[label=$\circ$]
	\item In the regime of weak gravitational deflections produced by local potential fluctuations, the orbital diffusion rate of \DM\@ particles was computed through a three-step procedure: (i)~quantifying the effect of single deflections, (ii)~characterising their cumulative statistical influence via the local power spectrum of potential fluctuations, and (iii)~phase-averaging this spectrum along the orbits to obtain diffusion coefficients as functions of orbital parameters.
	\item This formalism was applied to the cusp-to-core transformation of \DM\@ haloes induced by the fluctuating gravitational potential of repeated supernova explosions.
	\item Analytical expressions for the corresponding relaxation rates were derived for radial and general orbits, and were shown to reproduce accurately the results of dedicated Monte Carlo simulations in the regime where the approximations hold.
	\item The rate of cusp flattening was then obtained from the corresponding \FP\@ equations for isotropic  \DFs\@.
\end{itemize}
More specifically, the key results of this work are given by equation~\eqref{eq:ScalingRelation}, which gives  the extension of the diffusion coefficient to general orbits, introducing a dependence on angular momentum that suppresses diffusion for more circular trajectories; 
equation~\eqref{eq:DEEofEScaling}, which gives the scaling relation for the energy diffusion coefficient of radial orbits, capturing its dependence on energy and the location and amplitude of maximum diffusion;
and 
equations~\eqref{eq:tD} and \eqref{eq:tDwithNumbers} which give the expression for the characteristic diffusion timescale, relating it to the physical properties of \SBs\@ and the host halo.
\figref{fig:DEEofE} shows the energy diffusion coefficient for radial orbits, illustrating its bell-shaped dependence on energy and the agreement between theory, scaling relations, and simulations.
\figref{fig:DTheory3D} shows the orbit-averaged diffusion coefficient in $(E,L)$ space, demonstrating that diffusion decreases with increasing angular momentum at fixed energy.
Finally, \figref{fig:FPEvolution} shows the time evolution of the inner density slope, illustrating the progressive flattening of the cusp,  in agreement with theoretical expectations.

\subsection{Perspective}\label{subsec:Perspective}
The generality of the strategy presented in this paper makes it transposable to other baryonic feedback-driven secular processes, including the dynamical heating of stellar discs or the gradual dissolution of galactic bars by giant molecular clouds, supernovae, or interstellar turbulence. Relaxation times could also be potentially computed conditional on a given environment. More generally still, it could capture jointly and parametrically the long-term orbital impact of alternative models of \DM\@. In that context, it should allow astronomers to disentangle the effects of baryons, so as to put quantitative constraints on these models. Let us briefly discuss these extensions.
\begin{itemize}[label=$\circ$]
\item \textit{Vertical heating and radial migration} Beyond the so-called cusp-to-core transformation, other dynamical heating processes can be modeled analytically in this orbit-averaging framework, by changing the target \DF\@ and perturbation sources. For example, vertical disc heating is identified as the likely process for secular formation of thick discs \citep[e.g.,][]{Park2021,Tsukui2025,Oh2026}. In that context, equation~\eqref{eq:DEEofEScaling} can be transposed to capture the rate of change of vertical oscillation, instead of radial ones. Along the same line, it is worth noting that deflections by \GMCs\@ \citepalias[e.g.,][]{BinneyLacey1988} fall within the present framework via a flip in sign, reflecting that as they form, \GMCs\@ are transients which swallow a given region of gas before condensing into a star-forming region. In this context, it would be of interest to account for the cooler dynamical temperature of the disc, which will boost the orbital diffusion quadratically in the gravitational susceptibility of the disc \citep{Roule2025}. The framework also applies to radial migration \citep[e.g.,][]{SellwoodBinney2002,Frankel2020,Modak2026Interstellar}.
\item \textit{Bar dissolution} The dissolution of galactic bars is another process which could be driven by baryonic fluctuations \citep[e.g.,][]{Weinberg2025}. In the co-rotating frame of the bar, the orbital diffusion of radial orbits of a given Jacobi integral, $E_{J}$ (e.g., \S{3.3.2} from \citetalias{BinneyTremaine2008}), would also be captured by a variation of equation~\eqref{eq:DEEofEScaling}.
\item \textit{Alternative models of \DM\@}
  The effect of alternative \DM\@ (e.g., fuzzy) can also be modeled effectively as external perturbations from transient quasi-particles \citep[e.g.,][]{BarOr2019,ElZant2020,BarOr2021} whose size and lifetime, set by the de Broglie wavelength, are direct analogs of $\Rbub$ and $\Tbub$. On the other hand, self-interacting \DM\@ instead adds a local, short-range collision operator to the kinetic equation, alongside the long-range gravitational one, so that its effects are described by an effective scattering cross-section \citep[e.g.,][]{Balberg2002,KodaShapiro2011,Kummer2018} which could be added to baryonic feedback-driven diffusion.
\item \textit{Diffusion from multiple sources} 
As long as the deflections can be treated in the perturbative regime, we can replace the potential fluctuation in equation~\eqref{eq:EquationOfMotion} with the total fluctuations, arising from multiple sources (SNe, GMCs, turbulence, alternative \DM\@ model). If the correlations between perturbers are negligible (e.g., due to non-overlapping frequency ranges, or weak mutual interactions), then the total power spectrum becomes a linear combination of individual power spectra. Generally, observables arising from a combination of multiple processes are subject to partial degeneracies, which could be lifted (i)~provided their $\calEmax$ are sufficiently different and (ii)~via the joint fit of observables encoding the various contributions differently (e.g., steepness of rotation curve, anisotropy profile etc.). 
\item \textit{Cosmic environments} 
At leading order, the impact of large-scale environments (e.g., clusters or the cosmic web) on the morphology of galaxies could eventually be addressed by extending the result of this paper to capture the joint impact of deflectors within the galaxy and the cluster, and accounting for the convoluted intra-galactic and intra-cluster trajectories \citep{Yoo2024,Jeon2026}. After marginalisation over orbital configurations, one could compute galactic relaxation rates conditional on the binding energy of the galaxy within the cluster (e.g., Appendix D in \citealt{Hong2024}). 
\item \textit{From non-resonant to resonant theory} A resonant approach \citep{PichonAubert2006} will be required to explain the deviations beyond local (small size, shortly lived \SBs\@)  and uncorrelated sources (massless test \DM\@ particle and independently distributed, background \SB\@). This can be done  at the expense of
the flexibility provided by phase averaging, which decouples the microscopic process from the global angle actions of the galaxy  \citep{Theuns1996,Fouvry2021,Tep2026}. 
	
\end{itemize}
\section*{Data availability}

The data underlying this article are available upon reasonable request to the authors. The code {\textsf{OrbitAveragedFokkerPlanck}}, written in {\textsf{Julia}}~\citep{JuliaCite}, is available at the URL: \href{https://github.com/eunheeko/OrbitAveragedFokkerPlanck.git}{https://github.com/eunheeko/OrbitAveragedFokkerPlanck.git}.

\section*{Acknowledgements}
We warmly thank M. Weinberg, J. Binney, E. Vasiliev, R. Ewart, J. Ko, J. Yoo, J. Shin, J. Kim and M. Roule for valuable comments during the various stages of this project.   
This work is partially supported by the grants 
GALBAR ANR-25-CE31-4684 and BEYOND-BL ANR-25-CE57-2626 of the French Agence Nationale de la Recherche. 
This project has received financial support from the CNRS through the MITI interdisciplinary programs.
This work has made use of the Infinity Cluster hosted by Institut d'Astrophysique de Paris, partially funded by IDF-DIM-ORIGINES-2023-4-11. 
We thank St\'ephane Rouberol for the smooth running of the Infinity cluster. Artificial intelligence was  used in places to refine the phrasing and improve readability.
We thank Campus France and the {\it Ministère de l'Europe et des Affaires étrangères} for funding this research. This research was supported in part by grant NSF PHY-2309135 to the Kavli Institute for Theoretical Physics (KITP).

\appendix
\section{Frame and coordinates}\label{app:sec:FrameAndCoordinates}
\begin{figure*}[!htbp]
	\centering
\begin{subfigure}[b]{0.45\textwidth}
		\centering
		\includegraphics[width=0.90\textwidth]{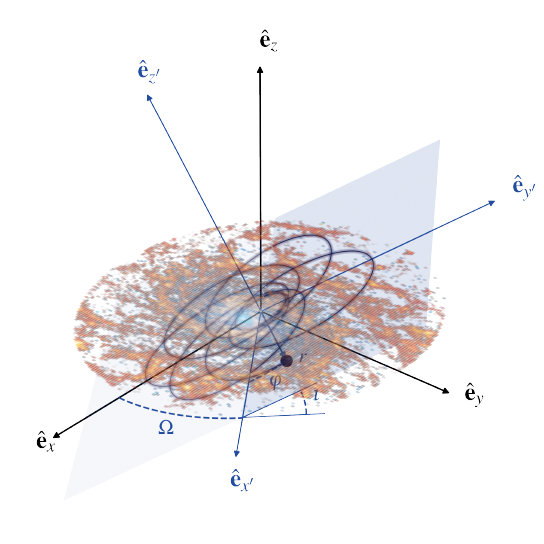}
		\caption{ Position frame }
	\end{subfigure}
\begin{subfigure}[b]{0.45\textwidth}
		\centering
		\includegraphics[width=0.90\textwidth]{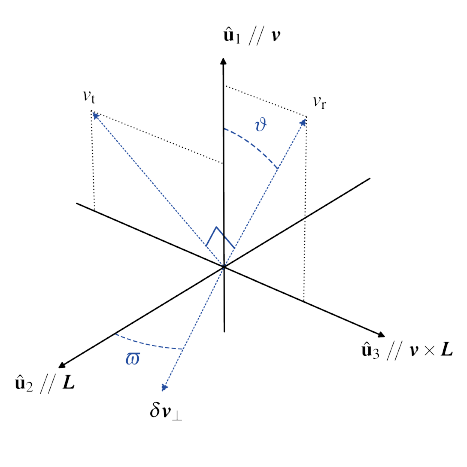}
		\caption{ Velocity frame}
	\end{subfigure}
\caption{ 
	Left panel (a): a reference (unprimed) frame with unit basis vector $(\uv{x}, \uv{y}, \uv{z})$ and a halo orbital frame with $(\uv{x^{\prime}}, \uv{y^{\prime}}, \uv{z^{\prime}})$ to describe the position of a test particle. The orbital frame is rotated relative to the reference frame by the longitude of the ascending node $\Omega$ and inclination angle $\iota$. In the primed frame, the \DM\@ particle is located at $r$ from the centre with orbital phase $\varphi$. Right panel (b): a Cartesian velocity frame: $(\uvu{1}, \uvu{2}, \uvu{3})$ comoving with the central position of \SB\@, where $\uvu{1} \! \parallel \! \bv$, $\uvu{2} \! \parallel \! \bL$, and $\uvu{3} \! \parallel \! \bv \times \bL$. The radial velocity vector $\vr$ and the tangential velocity $\vt$ lie in the ${(\uvu{1}, \uvu{3})}$ plane, making an angle $\vartheta$ between $\vr$ and $\uvu{1}$ axis. The local velocity deflection perpendicular to the initial velocity $\delta \bv_{\perp}$ is on the ${(\uvu{2}, \uvu{3})}$ plane, with azimuthal angle $\varpi$.
	}
	\label{fig:Frame}
\end{figure*}
In this appendix, we define a coordinate frame of the test particle's position and velocity, illustrated in  \figref{fig:Frame}.

Let $\br = (x, y, z)$ denote the position vector of a test particle on a spherical halo orbit, expressed in the reference frame (unprimed frame), whose $(\uv{x}, \uv{y})$ plane aligns with the stellar disc (see the left panel of \figref{fig:Frame}). We also introduce an orbital frame (the primed frame), whose $(\uv{x^{\prime}}, \uv{y^{\prime}} )$ plane coincides with the orbital plane of the test particle. For an orbit of radius $r$ and orbital phase $\varphi$, the position vector in the halo frame is given by $(x^{\prime}, y^{\prime}, z^{\prime}) = (r\cos{\varphi}, r\sin{\varphi}, 0)$. 

The orbital plane is inclined relative to the stellar disc plane by an inclination angle $\iota$, and a longitude of the ascending node $\Omega$. Accordingly, the position coordinate in the unprimed frame is given by 
\begin{subequations}\label{eq:XYZ}
\begin{align}
	x &= r\cos{\varphi}\cos{\Omega} - r\sin{\varphi}\cos{\iota}\sin{\Omega} \, , \\
	y &= r \cos{\varphi} \sin{\Omega} + r \sin{\varphi} \cos{\iota} \cos{\Omega} \, , \\
	z &= r \sin{\varphi} \sin{\iota} \, .
\end{align}
\end{subequations}
The transformation relation between the basis of the reference frame and the orbital frame is then given by
\begin{align}
\begin{bmatrix}
\uv{x^{\prime}} \\ \uv{y^{\prime}} \\ \uv{z^{\prime}}
\end{bmatrix}
&= 
\begin{bmatrix}
\cos{\Omega} & \sin{\Omega} & 0 \\
-\cos{\iota} \sin{\Omega} &  \cos{\iota} \cos{\Omega} & \sin{\iota} \\
\sin{\iota} \sin{\Omega} & -\sin{\iota} \cos{\Omega} & \cos{\iota}
\end{bmatrix}
\begin{bmatrix}
\uv{x} \\ \uv{y} \\ \uv{z}
\end{bmatrix} \, .
\end{align}

We introduce a Cartesian velocity frame to compute the local diffusion coefficients, as illustrated in the right panel of \figref{fig:Frame}. The unit basis vectors $(\uvu{1}, \uvu{2}, \uvu{3})$ are aligned with the \DM\@ velocity $\bv$, the angular momentum vector $\bL$, and $\bv \times \bL$, respectively. Denoting the angle between the position vector $\br$ and $\uvu{1}$ by $\vartheta$, and the orientation angle of the velocity perturbation perpendicular to the motion, $\delta \bv_{\perp}$, relative to $\uvu{2}$ by $\varpi$, the vectors $(\br, \bv, \bL)$ are expressed as
\begin{subequations}\label{eq:rbLinComovingFrame}
\begin{align}
	\br &= r \cos \vartheta \, \uvu{1} + r \sin \vartheta \, \uvu{3} \, , \\
	\bv &= v \, \uvu{1} \, , \\
	\bL &= L \, \uvu{2} \, .
\end{align}
\end{subequations}
\section{Local diffusion coefficients}\label{app:sec:LocalDiffusionCoefficients}
To obtain the orbit-averaged diffusion coefficients in terms of integrals of motion such as $\calE$ and $L$, we must (i)~calculate the local velocity-space diffusion coefficients as specified in equation~\eqref{eq:LocalDiffusionCoefficientInVelocity}, and (ii)~transform them into diffusion coefficients in energy and angular momentum space. 
\subsection{Local deflection in velocity}\label{app:subsec:LocalDeflectionInVelocity}
In general, the local velocity deflection by a single \SB\@ at fixed $\br$  at a given time  can be decomposed  into components parallel ($\delta \vparl$, denoted by $\parallel$) and perpendicular ($\delta \vperp$, denoted by $\perp$) to the incoming velocity as
\begin{align}\label{eq:deltabv1}
	\delta \bv &= \delta \vparl \, \uvu{\parallel} + \delta \vperp \uvu{\perp} \, .
\end{align}
In the $(\uvu{1}, \uvu{2}, \uvu{3})$ frame defined in \appref{app:sec:FrameAndCoordinates}, equation~\eqref{eq:deltabv1} takes the form
\begin{subequations}\label{eq:deltabv2}
\begin{align}
	\delta \vparl \, \uvu{\parallel} &= \delta v_{1} \, \uvu{1} \, , \\ \addlinespace
	\delta \vperp \uvu{\perp} 
	&= \delta \vperp \cos \varpi \, \uvu{2} + \delta \vperp \sin \varpi \, \uvu{3} \notag \\
	&= \delta v_{2} \, \uvu{2} + \delta v_{3} \, \uvu{3} \, ,
\end{align}
\end{subequations}
where $\varpi$ is an azimuthal angle in the ${(\uvu{2}, \uvu{3})}$ plane. Here, $\delta \bv$ is a stochastic quantity, whose randomness is encoded in $\delta \vparl$, $\delta \vperp$, and $\varpi$. Assuming a locally homogeneous distribution of \SBs\@, $\varpi$ is uniformly distributed over ${0 \leq \varpi \leq 2\pi}$, and independent of the random variable ${\delta v}$. 

From equations~\eqref{eq:deltabv2}, ensemble-averaging the first and second moments of velocity deflections over random deflection and $\varpi$, denoted by brackets, gives
\begin{subequations}\label{eq:EnsembleAverageddelvdelv}
\begin{align}
	\langle \delta v_{i} \rangle &=
	\begin{dcases}
		\langle \delta \vparl \rangle, \quad{} & \text{if }  i = 1 \\
		0, \quad{} &\text{if }  i = 2, 3
	\end{dcases}\\ \addlinespace
	\langle \delta v_{i} \, \delta v_{j} \rangle  &=
	\begin{dcases}
		\langle (\delta \vparl)^{2} \rangle, \quad{} &\text{if }  (i, j) = (1, 1) \\
		\frac{1}{2} \langle (\delta \vperp)^{2} \rangle, \quad{} &\text{if } (i,j) = (2, 2), (3,3) \\
		0 \quad{} &\text{if } i \neq j ,
	\end{dcases}
\end{align}
\end{subequations}
where  ${\langle \cos{\varpi} \rangle = \langle \sin{\varpi} \rangle = 0}$, while ${\langle \cos^{2}{\varpi} \rangle = \langle \sin^{2}{\varpi} \rangle \! =  1/2}$. 

Substituting equation~\eqref{eq:EnsembleAverageddelvdelv} into the definition of local diffusion coefficients given by equation~\eqref{eq:LocalDiffusionCoefficientDefinition} yields the non-zero velocity-space diffusion coefficients, decomposed into parallel and perpendicular components:
\begin{subequations}\label{eq:dvvFromLocal}
\begin{align}
	d_{\vparl} &= \lim_{T \to \infty} \frac{\langle \delta v_{1} \rangle}{T} \, , \\
	d_{\vparl \vparl} &= \lim_{T \to \infty} \frac{\langle (\delta v_{1})^{2} \rangle}{T} \, , \\
	d_{\vperp \vperp} &= \lim_{T \to \infty} \frac{\langle (\delta v_{2})^{2} \rangle + \langle (\delta v_{3})^{2} \rangle}{T} \, .
\end{align}
\end{subequations}
Expressed in this basis, the first-order velocity drag vector $d_{\bv}$ and second-order velocity diffusion tensor $d_{\bv \bv}$ take the form
\begin{subequations}\label{eq:dvvTensor}
\begin{align}
	d_{\bv} &= d_{\vparl} \, \uvu{1} \, , \\
	d_{\bv \bv} &= d_{\vparl \vparl} \, \uvu{1} \otimes \uvu{1} + \frac{1}{2} d_{\vperp \vperp} \, (\mathbb{I} - \uvu{1} \otimes \uvu{1}) \, ,
\end{align}
\end{subequations}
where $\mathbb{I}$ is a ${3\times3}$ identity matrix.

\subsection{Local deflection in $(E, L)$}\label{app:subsec:LocalDeflectionInEL}
In terms of $(r, v, L)$, local deflection in $(E, L)$ up to second order in $\delta v$ can be written as 
\begin{subequations}\label{eq:delEL}
\begin{align}
	\delta E &=\frac{1}{2} (\bv + \delta\bv)^{2} - \frac{1}{2} \bv^{2} \notag \\
	             &= v \, \delta \vparl + \frac{1}{2}  \left[ (\delta \vparl)^{2} + (\delta \vperp)^{2} \right] + \calO(\| \delta \bv \|^{3}), \\ \addlinespace
	\delta L &=  \| \bL + \delta \bL \| - \| \bL\|  \\ 
	&= \frac{\bL \cdot \delta \bL}{L}+\frac{1}{2L}(\delta \bL)^{2} -\frac{1}{2L} \left(\frac{\bL \cdot \delta \bL}{L}\right)^{2} + \calO(\| \delta \bL \|^{3}) \notag \\
	&= r \delta \vparl \frac{\vt}{v} - r \delta \vperp \sin{\varpi} \frac{\vr}{v} + \frac{1}{2L} r^{2} (\delta \vperp)^{2} \cos^{2}{\varpi} + \calO(\| \delta \bv \|^{3}) \, , \notag
\end{align}
\end{subequations}
where we use ${\delta \bL = \br \times \delta \bv}$.
By ensemble-averaging equation~\eqref{eq:delEL} and using equation~\eqref{eq:dvvFromLocal}, the local diffusion coefficients in $(E, L)$ become
\begin{subequations}\label{eq:LocalDiffusionCoefficientInEL}
\begin{align}
d_{E} &= v \, d_{\vparl} + \frac{1}{2} (d_{\vparl \vparl} + d_{\vperp \vperp}) \, , \label{eq:dE}\\
d_{L} &=  \frac{L}{v} \, d_{\vparl} + \frac{r^{2}}{4L} \, d_{\vperp \vperp} \, , \label{eq:dL}\\
d_{EE} &= v^{2} \, d_{\vparl \vparl} \, , \\
d_{LL}
	&= \frac{L^{2}}{v^{2}} \, d_{\vparl \vparl} + \frac{1}{2} (r^{2} - \frac{L^{2}}{v^{2}}) \, d_{\vperp \vperp} \, , \\
d_{E L} &= L \, d_{\vparl \vparl} \, .
\end{align}
\end{subequations}
Equations~\eqref{eq:LocalDiffusionCoefficientInEL} are consistent with \S{C} in \citet{Bar-Or2016} and \S{F1} in \citet{Tep2022}.
\section{Modeling superbubbles}\label{app:sec:ModellingSuperbubbles}
\subsection{Evolution of superbubble}\label{app:subsec:EvolutoinOfSuperbubbles}
\begin{figure}[!htbp]
	\center\includegraphics[width = 0.48\textwidth]{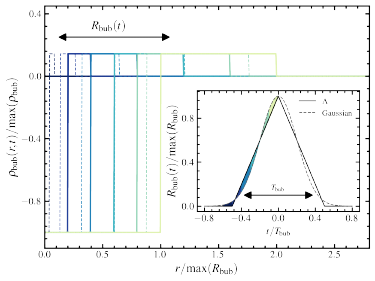}
	\caption{
	Spatiotemporal evolution of the \SB\@ mass density $\rhobub(r, t)$, defined in equation~\eqref{eq:SBMassDensity}. The profile features an underdense central cavity surrounded by an overdense outer shell. The characteristic size of the \SB\@ expands and contracts linearly in time (color-coded) to model the lifecycle of the density fluctuation (solid lines). Corresponding Gaussian approximations are depicted as dashed lines with matching colors. The temporal evolution of the \SB\@ radius for different profile choices is shown in the inset. This fluctuation profile can be adapted to accommodate various observational constraints, with the chosen profile determining the scaling of the resulting diffusion coefficients (see \appref{app:sec:ScalingRelation}).	}	
	\label{fig:Superbubble}
\end{figure}

Let us construct the power spectrum of \SBs\@ from the evolution of a \SB\@ density profile, illustrated in \figref{fig:Superbubble}. In the orbit-averaging framework, the fast relative motion between \DM\@ particles and \SBs\@, along with the sufficiently short ballistic timescale $\tbal$, implies that the orbit does not resolve the detailed internal structures of the \SBs\@ during encounters. Therefore, we focus on the dynamical aspect of \SBs\@, while simplifying the detailed structures that emerge from hydrodynamical processes \citep[e.g.,][]{MacLow1988,Keller2014,Yadav2017}.

Following a series of supernova explosions, the injected energy drives the expansion of the ambient medium, which subsequently relaxes and smooths out the density fluctuations over time. Motivated by this physical picture, we model the mass density profile as an inner underdense cavity ($\rhobubin < 0$) within radius $\Rbubin$, surrounded by an outer overdense shell ($\rhobubout > 0$) extending to $\Rbubout$:
\begin{align}\label{eq:SBMassDensity}
	\delta \rhobub(r, t; \Rbub, \Tbub) = 
	\begin{dcases}
		\rhobubin \, , \quad &\text{if } r \le \Rbubin(t) \, , \\
		\rhobubout \, , \quad &\text{if } \Rbubin(t) < r \le \Rbubout(t) \, , \\
		0 \, , \quad &\text{if } \Rbubout(t) < r \, .
	\end{dcases}
\end{align}
Here, the characteristic radius ${\Rbubin(t; \Tbub) = \Rbub \, \Lambda(t / \Tbub)}$ evolves in time over a characteristic lifespan $\Tbub$, where $\Lambda$ denotes a triangular window function (see the solid black line in the inset of \figref{fig:Superbubble}). The initial growth and subsequent decay of the width of  $\Lambda$ mimic the shock-driven expansion of the \SB\@ followed by the gravitational and thermal relaxation of the gas. 

Throughout this work, we set the outer \SB\@ radius to twice its inner radius, ${\Rbubout = 2 \Rbubin}$, following the work of \citet{Kipper2025}. Mass conservation then constrains the density of the outer overdense shell, such that ${-\rhobubin \, \Rbubin^{3} = \rhobubout \, (\Rbubout^{3} - \Rbubin^{3}) }$. 
See \appref{app:subsec:Superbubble} for the physical values adopted and their motivation. We find that modification of the ratio $\Rbubout/\Rbubin$ by a factor of order unity does not appreciably affect our numerical results or analytic approximations, when using different approximation schemes considered in \appref{app:sec:ScalingRelation}.

Locally, \SBs\@ are distributed homogeneously, while the $i^\mathrm{th}$ \SB\@ out of a total $\Nbub$ appears at a random position $\br_{i}$ and birth time $t_{i}$ uniformly (see \secref{subsec:LocalAveraging}). Note that $\Nbub$ is a function of a test particle's position, determined by the perturber's global distribution such as equation~\eqref{eq:SigmabubInR}. Then the density fluctuation, experienced by \DM\@ at fixed position $\br$, can be generalized into
\begin{equation}\label{eq:NSBMassDensity}
	\delta \rho(\br, t) = \sum_{i}^{\Nbub} \delta \rhobub(\br - \br_{i}, t - t_{i}; \Rbub, \Tbub) \, .
\end{equation}
Note that stochasticity enters only through the random variables $\br_{i}$ and $t_{i}$ within the density profile. 

Let us adopt the following Fourier transforms in space $\br$ (resp. time $t$), denoted by a subscript spatial frequency $\bk$ (resp. a tilde with temporal frequency $\omega$)
\begin{subequations}\label{eq:FourierTransform} \begin{align}
	\!\!\!\! \delta \Phi_{\bk}  &= \int \! \dd^{3} \br \, \delta \Phi(\br) \, \re^{-\ri \bk\cdot\br} \, ; \!\!\!&\delta \Phi(\br) &= \int \!\! \frac{\dd^{3} \bk}{(2\pi)^{3}} \, \delta \Phi_{\bk} \, \re^{\ri\bk\cdot \br} \, ,\\ \!\!\!\! \delta \tilde{\Phi} &= \int \! \dd t \, \delta \Phi(t) \, \re^{\ri \omega t} \, ;  \!\!\!&
	\delta \Phi(t) &= \int \! \frac{\dd \omega}{2\pi} \, \delta \tilde{\Phi} \, \re^{-\ri \omega t} \, .
\end{align}
\end{subequations}
Performing Fourier transform of equation~\eqref{eq:NSBMassDensity} in position and time factors this dependence out explicitly as
\begin{align}\label{eq:delrhokw}
	\delta \tilde{\rho}_{\bk}(\omega) 
	&= \sum_{i}^{\Nbub} \int_{-\infty}^{\infty} \!\! \dd t \int \!\! \dd^{3} \br \, \delta \rhobub( \, \abs{\br - \br_{i}}, t - t_{i}) \, \re^{-\ri (\bk \cdot \br - \omega t)} \notag \\
	&= \sum_{i}^{\Nbub} \int_{-\infty}^{\infty} \!\! \dd t^{\prime} \! \int \!\! \dd^{3} \br^{\prime} \delta \rhobub( \, |\br^{\prime}|, t^{\prime}) \, \re^{-\ri \bk \cdot (\br^{\prime} + \br_{i}) + \ri  \omega (t^{\prime} + t_{i})} \notag \\
	&=  \sum_{i}^{\Nbub} \delta \tilde{\rho}_{\text{bub}, \bk}(\omega) \, \re^{-\ri (\bk \cdot \br_{i} - \omega t_{i})} \, ,
\end{align}
where we change the variable by using ${\br^{\prime} = \br - \br_{i}}$ and ${t^{\prime} = t - t_{i}}$ in the second line. 

Using the property ${\delta \tilde{\rho}_{\text{bub}, \bk}^{*} = \delta \tilde{\rho}_{\text{bub}, -\bk}}$ for a real-valued $\delta \rhobub$, we compute the density power spectrum by ensemble-averaging the square of equation~\eqref{eq:delrhokw} over \SB\@'s random positions of $\br_{i}$ and $\br_{j}$, and random birth time of $t_{i}$ and $t_{j}$, denoted by bracket with subscripts $(i, j)$ as
\begin{align}\label{eq:delrhokw2}
	 \langle |\delta \tilde{\rho}_{\bk}(\omega)|^{2} \rangle 
	 &= \left \langle \sum_{i,j}^{\Nbub} \re^{-\ri (\bk \cdot \br_{i} - \omega t_{i}) + \ri (\bk \cdot \br_{j} - \omega t_{j})} 
	 |\delta \tilde{\rho}_{\text{bub}, \bk}(\omega)|^{2} \right \rangle_{i, j} \notag \\
	&=|\delta \tilde{\rho}_{\text{bub}}(\bk, \omega)|^{2} \,  \sum_{i,j}^{\Nbub} w_{ij} \, ,
\end{align}
where the ensemble average acts only on the exponential terms, and we define the weight $w_{ij}$ as
\begin{align}\label{eq:wij}
	w_{ij} 
	&\equiv \left \langle  \, \re^{-\ri (\bk \cdot \br_{i} - \omega t_{i}) + \ri (\bk \cdot \br_{j} - \omega t_{j})} \right \rangle_{i, j} \notag \\
	&= \int \! \frac{\dd^{3} \br_{i}}{V} \int \! \frac{\dd^{3} \br_{j}}{V} \, \re^{-\ri \bk \cdot (\br_{i} - \br_{j})} \int \! \frac{\dd t_{i}}{T} \int \! \frac{\dd t_{j}}{T} \, \re^{\ri \omega(t_{i} - t_{j})} 	 \notag \\
	&=
	\begin{cases}
	  \displaystyle  1\, , & \text{if } i = j \\
	  \displaystyle \frac{(2\pi)^{3}}{V} \dirac^{(3)}(\bk) \frac{2\pi}{T} \dirac(\omega) \, , & \text{if } i \neq j \, .
	 \end{cases}
\end{align}
Here, $V$ and $T$ represent the spatial volume and temporal duration corresponding to the uniform probability distributions for the birth positions $\br_{i}$ and times $t_{i}$, respectively. 
\begin{figure}[!htbp]
	\center\includegraphics[width=0.49\textwidth]{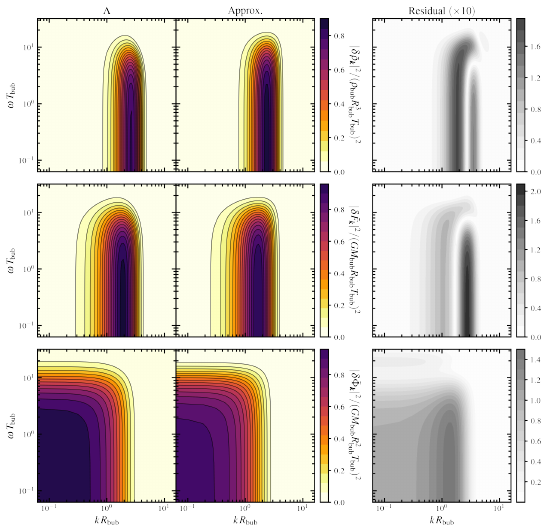}
	\caption{
	From top to bottom: the power spectra of density, force, and potential fluctuations as a function of $(k, \omega)$, scaled by corresponding normalization factors, respectively. The color codes indicate the amplitude of each spectrum. The left columns show the numerically integrated results from the exact model profile in equation~\eqref{eq:SBMassDensity}, while the middle columns display the corresponding analytic approximations. The right columns show the relative residuals between the two models, multiplied by a factor of $10$ for visual clarity. Overall, the analytic approximations reproduce both the shape and amplitude of the exact power spectra, with relative differences generally below $10\%$.
	}
	\label{fig:PowerSpectraRhoFPhi}
\end{figure}

Note that equation~\eqref{eq:wij} has two contributions from $i \!=\! j$ and $i \! \neq \!j$. However, the term from $i \! \neq \! j$ contributes only at $(\bk, \omega) = (\boldsymbol{0}, 0)$, and therefore does not affect the diffusion coefficient (see the resonance condition in equation~\ref{eq:LocalDiffusionCoefficientInVelocity}). Dropping the $i \! \neq \! j$ terms in the following, equation~\eqref{eq:delrhokw2} is simplified to
\begin{equation}
	 \langle |\delta \tilde{\rho}_{\bk}(\omega)|^{2} \rangle = \Nbub  |\delta \tilde{\rho}_{\text{bub}, \bk}(\omega)|^{2} \, .
\end{equation}

Using Poisson's equation ${\delta \Phi_{\bk} = -4 \pi G \delta \rho_{\bk} / k^{2}}$, we obtain the potential correlation as
\begin{equation}\label{eq:PotentialToDensityPS}
	 \langle |\delta \tilde{\Phi}_{\bk}(\omega)|^{2} \rangle = \Nbub \frac{(4 \pi G)^{2}}{k^{4}}  |\delta \tilde{\rho}_{\text{bub}, \bk}(\omega)|^{2} \, .
\end{equation}
In equation~\eqref{eq:LocalDiffusionCoefficientInVelocity}, the integration measure of $\bk$ can be written as  $\dd^{3}  \bk= \dd^{2} \bk_{\perp} \dd \bk_{\parallel} = 2\pi \kperp \dd \kperp \dd \kparl$ and hence the local diffusion coefficient reads
\begin{align}
d_{\bv \bv}  &=  \dotnbub \int_{0}^{\infty} \! \frac{\dd k_{\perp}}{2\pi} \int_{-\infty}^{\infty} \! \frac{\dd k_{\parallel}}{2\pi} k_{\perp} \, (\bk \otimes \bk) \, \tilde{\calP}_{\bk}(\omega = \kparl v) \, .
\end{align}
Note that ${d_{\vparl \vperp}}$ vanishes for a spherically symmetric \SB\@ power spectrum ${\tilde{\calP}_{k}}$, since the integrand is odd in $\kparl$. 
\subsection{Power spectra}\label{app:subsec:PowerSpectra}
To achieve analytic integration of the local diffusion coefficient, let us now construct an approximate power spectrum with a smoother integrable functional form. The sharp features in the exact form arise from the piece-wise shape of the mass density profile in equation~\eqref{eq:SBMassDensity}, which in turn leads to a highly oscillatory behaviour on small scales (high $k$). 

To circumvent this, we compute the Fourier-transformed density $\delta  \tilde{\rho}_{\bk}$, using a Gaussian profile that matches the full width at half maximum of the original Heaviside lambda-like profile (gray dashed line in the inset of \figref{fig:Superbubble}). We then introduce a second approximation for $| \delta \tilde{\Phi}_{\bk} |^{2}$, using a Pad\'e approximant, expanded around small $\Rbub$.

The resulting power spectra of density, force, and potential fluctuations, compared between the original model and the approximation, are shown in \figref{fig:PowerSpectraRhoFPhi}.

\section{Orbit-averaging}\label{app:sec:OrbitAveraging}
\subsection{Ergodicity}
Recalling equation~\eqref{eq:Ergodicity}, the time average for a general integrable 3D system \citep{Tep2025} with ${\bJ = (J_{r}, J_{\phi}, J_{\theta}) = (J_{r}, L_{z}, L - |L_{z}|)}$, where ${L_{z} = \bL \cdot \uv{z}}$ is the $z$-component of angular momentum, can be expressed as an average over the corresponding conjugate angle variables ${\btheta = (\thetar, \thetaphi, \thetatheta)}$
\begin{align*}
	\int_{0}^{T} \! \frac{\dd t}{T} &\leftrightarrow \underbrace{ \oint_{0}^{2\pi} \! \frac{\dd \thetar}{2\pi}}_{\displaystyle \langle \cdot \rangle_{\thetar}}
	\, \underbrace{\oint_{0}^{2\pi} \! \frac{\dd \thetaphi}{2\pi}}_{\displaystyle \langle \cdot \rangle_{\thetaphi}} \, 
	\, \underbrace{\oint_{0}^{2\pi} \! \frac{\dd \thetatheta}{2\pi}}_{\displaystyle \langle \cdot \rangle_{\thetatheta}} \, ,
\end{align*}
where the orbit averages over the corresponding angle variables are denoted by the bracket with the respective subscripts. 

For a spherical potential, one may equivalently use the actions ${\bJ = (J_{r}, L, L_{z})}$ with conjugate angles ${\btheta = (\theta_{1}, \theta_{2}, \theta_{3})}$, where ${\theta_{1} = \thetar}$, ${\theta_{2} = \thetatheta}$ and ${\theta_{3} = \Omega}$, defined in \appref{app:sec:FrameAndCoordinates} (see \S{3.5.2} in \citetalias{BinneyTremaine2008}). In this case, the longitude of the ascending node $\Omega$ is a constant of motion, and the time average involves only two angles, $\theta_{1}$, and $\theta_{2}$, taken at fixed $(\bJ, \Omega)$: 
\begin{equation}\label{eq:ErgodicitySpherical}
	\int_{0}^{T} \! \frac{\dd t}{T} 
\leftrightarrow \underbrace{ \oint_{0}^{2\pi} \! \frac{\dd \theta_{1}}{2\pi}}_{\displaystyle \langle \cdot \rangle_{\theta_{1}}}
	\, \underbrace{\oint_{0}^{2\pi} \! \frac{\dd \theta_{2}}{2\pi}}_{\displaystyle \langle \cdot \rangle_{\theta_{2}}} 
\leftrightarrow \frac{\Omegar}{\pi} \!\! \int_{\rp}^{\ra} \! \frac{\dd r}{\vr} 
	 \, \int_{0}^{2\pi} \!\! \frac{\dd \varphi}{2\pi}\, ,
\end{equation}
where the conjugate angle average is re-written in terms of the configuration space variables $(r, \varphi)$. The last step uses the fact that ${\theta_{2} - \varphi}$ depends only on $r$ along the orbit, so the $\varphi$-average must be taken first, at fixed $r$.

For a spherically symmetric distribution of perturbers, such as that given by equation~\eqref{eq:SigmabubInR}, the orbit-averaging over $\varphi$ becomes trivial. For distributions with a different geometry, such as the axisymmetric distribution, an explicit averaging over the relevant angular variables is instead required (see \appref{app:sec:BeyondCoplanarity}).

Because the power spectra are built from quantities defined in terms of positions and velocities, this equivalence lets us carry out the orbit
average without an explicit transformation from $(\br, \bv)$ to $(\bJ, \btheta)$. The method therefore applies to general orbits, for which
action--angle variables are not known in closed form, and to any perturbation of interest.
\subsection{Radial orbit averaging}\label{app:subsec:RadialOrbitAveraging}
The $\thetar$-averaged diffusion coefficient in $q$ follows
\begin{equation}\label{eq:RadialOrbitAveraging}
	\langle d_{q} \rangle_{\theta_{r}} = \oint_{0}^{2\pi} \frac{\dd \thetar}{2\pi} \,  d_{q}  = \frac{\Omegar}{\pi} \int_{\rp}^{\ra} \! \frac{\dd r}{\vr} \, d_{q} \, .
\end{equation}
In equation~\eqref{eq:RadialOrbitAveraging}, using $r$ as the integration variable is practically not convenient, since $1/\vr$ diverges at $\rp$ and $\ra$, where $\vr \to 0$. Although this divergence is integrable, it can be avoided by using H\'enon anomaly \citep{Henon1971}. First, we forward map the initial orbital labels from $(E, L)$ (or equivalently the effective semi-major axis and eccentricity $(a, e)$; \citealt{LyndenBell2015}) to radial and azimuthal orbital frequency $(\Omegar, \Omegaphi)$, after the substitution below (see Appendix A.1 in \citealt{Petersen2024}). The frequencies read (see equations 28 and 29 in \citealt{TremaineWeinberg1984})
\begin{equation}\label{eq:FrequencyRatio}
	\frac{1}{\Omegar} = \frac{1}{\pi} \int_{\rp}^{\ra} \frac{\dd r}{\vr}\, ; \quad{} \frac{\Omegaphi}{\Omegar} = \frac{L}{\pi} \int_{\rp}^{\ra} \frac{\dd r}{r^{2} \vr} \, .
\end{equation}
Next, introducing the H\'enon anomaly $u \in [-1, 1]$ \citep{Henon1971}, 
\begin{equation}\label{eq:HenonAnomaly}
	r(u) = a \, [1 + e f(u)]\,; \quad{} f(u) = u \left( \frac{3}{2} - \frac{1}{2} u^{2} \right) \, ,
\end{equation}
and the integration variable
\begin{equation}\label{eq:ThetaFuncion}
	\Theta(u) \equiv \frac{1}{\vr} \frac{\dd r}{\dd u} \, ,
\end{equation}
equation~\eqref{eq:RadialOrbitAveraging} can be rewritten with a finite integrand as
\begin{equation}\label{eq:RadialOrbitAveraginginU}
	D_{X} = \frac{\Omegar}{\pi} \int_{-1}^{1} \! \dd u \, \Theta(u) \, d_{X} \, .
\end{equation}
To perform $\thetar$-averaging in equation~\eqref{eq:RadialOrbitAveraginginU}, we rely on the \textsf{OrbitalElements.jl} package \citep{Petersen2024}, which implements the transformation in \eqref{eq:FrequencyRatio}--\eqref{eq:ThetaFuncion} numerically. 

\section{NFW dark matter halo}\label{app:sec:NFW}
For a cuspy gravitational potential, we adopt the \NFW\@ profile, defined as
\begin{equation}\label{eq:NFWPotential}
	\Phi(r) = -\Vhalo^{2} \frac{\Rhalo}{r} \log{\left( 1 + \frac{r}{\Rhalo} \right)} \, ,
\end{equation}
where $\Vhalo$ is a characteristic circular velocity and $\Rhalo$ is a halo scale radius. The corresponding \NFW\@ density profile follows
\begin{align}\label{eq:NFWDensity}
	\rho(r) \!=\! \frac{\rhohalo}{\left( r/\Rhalo  \right) {\left(1\!+\! r/\Rhalo  \right)^{2}}}\,, \, & \text{where } \Vhalo^{2} \!=\! 4\pi G \rhohalo \Rhalo^{2} \, .
\end{align}
For convenience, let us define the following dimensionless variables 
\begin{align}
\xi \equiv 1 - E/\Vhalo^{2}\,; \,\,
\ell \equiv L / \Lc\,; \,\,
\psi \equiv 1- \Phi/\Vhalo^{2}\,; \\ \notag
\scrr \equiv r / \Rhalo \, ; \, \,
\scrf \equiv f / f_{0} \, ; \, \, \text{and } \varrho \equiv \rho / \rhohalo \, .
\end{align}
For an isotropic \DF\@, we adopt the fitting formula introduced by \citet{Widrow2000}, which reproduces the limiting behaviour as $\xi \to 0$ and $\xi \to 1$ (see Appendix therein),
\begin{equation}\label{eq:IsotropicDF}
	\fiso(\xi) = f_{0} \, \xi^{3/2} (1 - \xi)^{-\lambda} \left( \frac{-\ln{\xi}}{1 -\xi } \right)^{q} \re^{P} \, ,
\end{equation}
where the normalization is chosen  such that ${M = \int  \dd^{3} \br \, \dd^{3} \bv f}$ with  $M$ the total mass. The corresponding density of states $\giso(\xi)$ such that ${\dd M / \dd \xi = \fiso(\xi) \, \giso(\xi)}$ follows
\begin{equation}\label{eq:IsotropicDoS}
	\giso(\xi) = 16 \sqrt{2} \pi^{2} \int_{0}^{\scrrc} \dd \scrr \, \scrr^{2} \sqrt{\psi - \xi} \, ,
\end{equation}
where $\scrrc$ is a circular orbit radius of energy $\xi$. \section{Beyond the coplanarity assumption}\label{app:sec:BeyondCoplanarity}
\begin{figure*}[!htbp]
	\centering
	\includegraphics[width=0.9\textwidth]{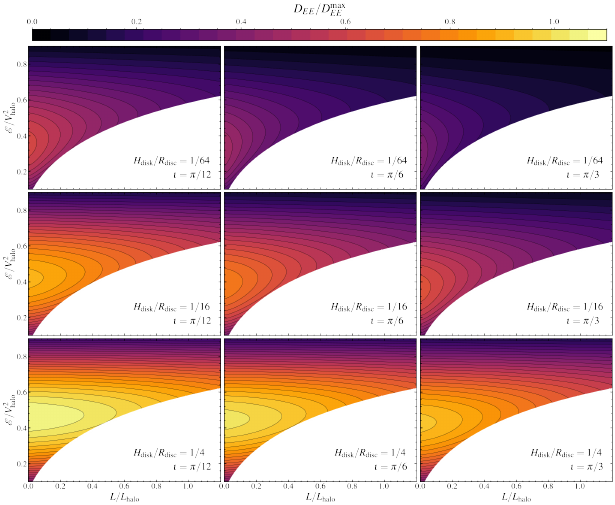}
	\caption{
Theoretical predictions of the orbit-averaged diffusion coefficients in energy, $D_{EE}$, scaled by their corresponding normalizing factors, for varying inclination angles $\iota = \pi/12, \pi/6$, and $\pi/3$ (left to right) and disc scale height ratios $\Hdisc/\Rdisc = 1/64, 1/16$, and $1/4$ (top to bottom). As the stellar disc becomes thinner (smaller $\Hdisc/\Rdisc$), $\calEmax$ shifts toward more tightly bound orbits, indicating that diffusion concentrates near the galactic centre and becomes relatively less efficient in the outskirts. A similar trend is observed at a fixed $\Hdisc/\Rdisc$ as the orbital inclination increases. These results demonstrate that the global spatial distribution of perturbing sources dictates the encounter rates experienced by \DM\@ particles, thereby shaping the structure of the diffusion coefficients across phase space.
	}
	\label{fig:DTheoryAxisymmetric}
\end{figure*}
In this appendix, we consider an alternative $\Sigmabub$ profile and examine its effect on the predicted orbit-averaged diffusion coefficients. As an extension of the spherically symmetric profile in equation~\eqref{eq:SigmabubInR}, we introduce an axisymmetric \SB\@ distribution, whose isocontours are concentric oblate spheroids, similar in geometry to those used in models of flattened galactic systems \citep[e.g.,][]{MiyamotoNagai1975}:
\begin{equation}\label{eq:SigmabubInRZ}
	\Sigmabub(r, z; \Rdisc, \Hdisc, \iota) = \frac{1}{1 + \sqrt{(r/\Rdisc)^{2} + (z/\Hdisc)^{2}}} \, ,
\end{equation}
where $\Rdisc$ and $\Hdisc$ denote the disc scale radius and height respectively, and $\iota$ is the inclination angle (see \appref{app:sec:FrameAndCoordinates}). Using the coordinate transformation in equation~\eqref{eq:XYZ}, equation~\eqref{eq:SigmabubInRZ} can be rewritten in terms of the orbital frame coordinate $(r, \varphi)$ as 
\begin{equation}\label{eq:SigmabubInRZ2}
	\Sigmabub(r, \varphi; \Rdisc, \Hdisc, \iota) = \frac{1}{1+r/\Rdisc^{\prime}} \, ,
\end{equation}
where the effective disc radius $\Rdisc^{\prime}$ is defined as
\begin{equation}\label{eq:ModifiedRdisc}
 \Rdisc^{\prime}\!\equiv\! \frac{\Rdisc}{\sqrt{\cos^{2}{\varphi} \!+\! \sin^{2}{\varphi} \left( \cos^{2}{\iota}\! +\! \frac{\Rdisc^{2}}{\Hdisc^{2}} \sin^{2}{\iota} \right)}} \, .
\end{equation}
Note that equation~\eqref{eq:ModifiedRdisc} depends on the orbital phase $\varphi$. In the spherical limit, ${\Rdisc = \Hdisc}$, or for halo orbits coplanar to the stellar disc as in the left panel of \figref{fig:Schematic}, ${\iota = 0}$, the original equation~\eqref{eq:SigmabubInR} is recovered.

When calculating the diffusion coefficients for a general spherical system of $\bJ = (J_{r}, L, L_{z})$ with equation~\eqref{eq:ModifiedRdisc}, the dependence of $\Rdisc^{\prime}$ on $\varphi$ requires an additional average over this angle. Thus, the orbit-averaging shown in equation~\eqref{eq:Ergodicity} must now be generalized to include the average over $\theta_{2}$ (see equation~\ref{eq:ErgodicitySpherical}). 

\figref{fig:DTheoryAxisymmetric} displays the resulting prediction of diffusion coefficients for different inclination angles and disc scale heights in equation~\eqref{eq:SigmabubInRZ2}. The remaining integral over $\varphi$ is evaluated numerically using the trapezoidal rule. This implies that different perturbation configurations alter the effective disc scale length and the consequent number of encountered sources. Orbit-averaging the local diffusion with the characteristic profile of $\Sigmabub$ therefore dictates the overall structure of the orbit-averaged diffusion coefficients.

For example, as the structure approaches a thinner disc configuration (decreasing $\Hdisc / \Rdisc$ ratio from the bottom to the top panels), the maximum diffusion peak shifts toward more tightly bound orbits (smaller $\calEmax$). A similar trend is observed when increasing the inclination angle (from the left to the right panels).

\section{Numerical simulations}\label{app:sec:NumericalSimulation}
For validation, let us simulate the evolution of \DM\@ trajectory and background \SBs\@ on a finite slab-like grid of size ${(16.0 \, \Rhalo ) \times (16.0 \, \Rhalo ) \times (0.25 \, \Rhalo)}$ for a duration of ${40 \, \thalo}$. We adopt a smaller vertical extent of the grid and place the orbital plane at the mid-plane ${z = 0}$. We adopt the physical units so that ${G = M = \Rhalo = 1}$.

\subsection{Initial orbital phases}\label{app:subsec:InitialOrbitalPhases}
The initial position and velocity $(\br_{0}, \bv_{0})$ of the \DM\@ particle are generated through rejection sampling. At a given orbital label $(E, L)$, the radial and tangential velocity $(\vr, \vt)$ are given by
\begin{align}
	\vr &= \pm \sqrt{2[E - \Phi(r)] -L^{2}/r^{2}} \, ; &
	\vt &= L / r \, ,
\end{align}
which defines the allowed radial range ${r \in [\rp, \ra]}$. Here, the pericentre $\rp$ and apocentre $\ra$ are the roots of $\vr = 0$, and consequently also determine the corresponding tangential velocity. 

Since the radial coordinate $r$ is not uniform in configuration space, we instead sample it using the H\'enon anomaly in equation~\eqref{eq:HenonAnomaly}, which provides uniform sampling in $u \in [-1, 1]$. We implement this using the {\textsf{JuliaStellarDynamics}} code \citep{Petersen2024}.
\subsection{Superbubbles}
The force fluctuations by \SBs\@ are generated independently of the \DM\@ orbits. Starting from the density profile in equation~\eqref{eq:SBMassDensity}, we solve Poisson's equation 
\begin{align}
	\frac{1}{r^{2}} \frac{\p}{\p r} \left( r^{2} \frac{\p \delta \Phi}{\p r} \right) = 4 \pi G \delta \rhobub \, ,
\end{align}
which yields the corresponding radial force profile of a single \SB\@ as a function of the distance from its centre as
\begin{align}\label{eq:SBForce}
	\!\! -\dfrac{\p \delta \Phi}{\p r} \!&=\! -\frac{4 \pi G}{r^{2}} \int_{0}^{r} \! \dd r^{\prime} \rhobub(r^{\prime}) \, r^{\prime \, 2} \notag \\
	\!\!\! &=\! -\frac{4 \pi G}{3}
	\begin{dcases}
	\rhobubin \, r \,, \!\! &\text{if } r \leq \Rbubin \\
	(\rhobubin-\rhobubout) \frac{\Rbubin^{3}}{r^{2}} + \rhobubout r \,,  \!\! &\text{if } \Rbubin < r \leq \Rbubout \\
	0 \, ,  \!\!&\text{if } \Rbubout < r \, .
	\end{dcases}
\end{align}

We then create the force fluctuation fields by adding the force given by equation~\eqref{eq:SBForce} to the background gravitational mean-force field for all the \SBs\@. The \SBs\@ are assigned random birth times uniformly over the simulation time $\Tsim$. Their spatial positions are sampled according to the radial distribution given by equation~\eqref{eq:SigmabubInR} with uniform azimuthal angle in $[0, 2\pi]$ and uniform vertical ($z$) distribution over $[-0.125\Rhalo, 0.125\Rhalo ]$. As each \DM\@ orbit always lies in the $z = 0$ plane, this axisymmetric slab-like configuration is intended to match the expected encounter rate $\dotnbub$, under the locally homogeneous but globally inhomogeneous distribution of \SBs\@.

Note that this prescription for the \SB\@ spatial distribution is not strictly consistent with the assumption that the system is treated as locally homogeneous but globally inhomogeneous (e.g., see \appref{app:subsec:EvolutoinOfSuperbubbles}). In practice, however, there is no sharp separation between these two scales, as discussed in \secref{subsec:Limitations}. Furthermore, $\Rbub$ should exceed the resolution limit to securely sample the frequency range of the power spectrum  (see \appref{app:secTruncationInFrequency}).
\subsection{Orbit integration and  sampling deflections}\label{app:subsec:OrbitIntegration}
We employ a leapfrog integration scheme (e.g., \S{3.4.2} of \citetalias{BinneyTremaine2008}) to evolve the system over a simulation time ${\Tsim = 40 \, \thalo}$ using a time step ${\Delta t = 10^{-4} \, \thalo}$. The numerical setup is identical for the \Mkv\@ and \nMkv\@ simulations, except for the applied forces: orbits in the \Mkv\@ simulation evolve exclusively under the mean-field \NFW\@ potential defined in equation~\eqref{eq:NFWPotential}, whereas the \nMkv\@ simulation incorporates both the mean-field potential and the \SB\@-induced fluctuating fields given by equation~\eqref{eq:SBForce}.

This distinction naturally leads to a different method for sampling deflections. In the \nMkv\@ simulation, the orbit changes its energy and angular momentum continuously, automatically retaining correlations from successive collisions. Conversely, in the \Mkv\@ simulation, local deflections in energy and angular momentum are sampled independently at each time step, following equation~\eqref{eq:delEL}. As a result, the orbit retains no memory of previous kicks. The cumulative effect of these local deflections is applied a posteriori to emulate a diffusion process consistent with the \Mkv\@ assumption.

\subsection{Measurement: ballistic vs. diffusive}\label{app:subsec:MeasurementOfD}
\begin{figure}[!htbp]
	\centering\includegraphics[width = 0.48 \textwidth]{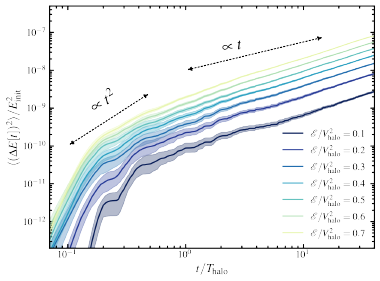}
	\caption{Evolution of the energy variance normalized by the initial energy $E_{\mathrm{init}}$, ${\langle (\Delta E[t])^{2} \rangle / E_{\mathrm{init}}^{2}}$, as a function of time, averaged over $128$ realizations of \SB\@ distributions and $1\,024$ initial orbital conditions from the \Mkv\@ numerical simulations. Different colors represent different initial values of $\calE$. This figure illustrates the transition from the ballistic to the diffusive regime. The shaded bands indicate the $16^{\mathrm{th}}$--$84^{\mathrm{th}}$ percentile bootstrap intervals for the mean, obtained by resampling the ${n = 128}$ \SB\@ realizations after averaging over the initial orbital conditions. To accurately extract the diffusion coefficients, the sampling time window must be restricted to the linear diffusive regime.
	}
	\label{fig:DiffusionMeasurement}
\end{figure}
In order to measure the diffusion coefficient in simulations, one should (i)~track the deviation in the variables of interest as a function of time, (ii)~average the deviations over different realizations of orbital phases and perturbations, and (iii)~determine their slope in the diffusive regime, where the evolution becomes linear in time. 

Within the ballistic regime, correlations between successive perturbations are non-negligible, causing the variance of the fluctuations to grow quadratically with time. Therefore, the system must evolve long enough for these correlations to decay, allowing it to enter the diffusive regime where the variance grows linearly in time \citep[e.g.,][]{Risken1989,Roule2022}:
\begin{equation}
\int_{0}^{T} \! \dd t \int_{0}^{T} \! \dd t^{\prime} \, \calP_{\bk}(t - t^{\prime}) \, \re^{\ri \bk \cdot \bv (t - t^{\prime})} 
\propto  
\begin{dcases}
\displaystyle T^2 \, , & \text{if } T \ll \tbal \\
\displaystyle \tbal \, T \, , & \text{if } T\gg \tbal \, .
\end{dcases}
\end{equation}

Even in this diffusive regime linear in time, one should pick a time lag during which the physical quantity ($E$ or $L$ here) should not vary much (small enough lag), but the \Mkv\@ assumption is satisfied (large enough lag) \citep{vanKampen1992,Hamers2014}.

An example of this behaviour is shown in \figref{fig:DiffusionMeasurement}. We discard the initial time window, $t/\thalo \lesssim 1$, which corresponds to an elapsed time shorter than both the dynamical timescale, $\tdyn$, and the perturbation timescale, $\Tbub$, following the onset of the simulation.
\subsection{First-order diffusion coefficients}\label{app:subsec:DTheoryFirstOrder}
\begin{figure}[!htbp]
	\centering
	\begin{subfigure}[b]{0.50\textwidth}
		\centering
		\caption*{(a) $D_{E}$}
		\includegraphics[width=1.0\textwidth]{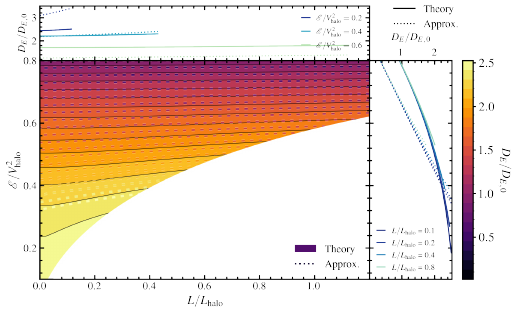}
	\end{subfigure}
	\begin{subfigure}[b]{0.50\textwidth}
		\centering
		\caption*{(b) $D_{L}$}
		\includegraphics[width=1.0\textwidth]{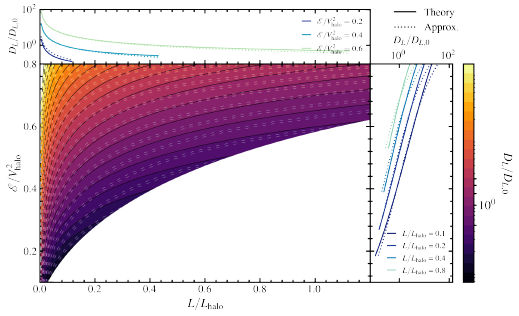}
	\end{subfigure}
	\caption{
Theoretical prediction of orbit-averaged diffusion coefficients $D_{E}$  (top panel) and $D_{L}$ (bottom panel), scaled by the normalization factors defined in equation~\eqref{eq:D0sets2}, for the same setup as in \figref{fig:DTheory3D}. Line contours depict the approximate scalings given by equation~\eqref{eq:ScalingRelationFirstOrder}. As in \figref{fig:DTheory3D}, the approximate scaling relations agree well with the theoretical prediction in both amplitude and shape.
	}
	\label{fig:DTheoryFirstOrder}
\end{figure}

We do not present numerical measurements of the first-order drift coefficients, $D_E$ and $D_L$, as their ensemble-averaged slopes cannot be reliably extracted from highly oscillatory realizations \citep{Theuns1996,Hamers2014,Merritt2015}. Because first-order deflections scale linearly with the perturbations rather than quadratically (see equations~\ref{eq:dE}--\ref{eq:dL}), they exhibit large-amplitude, zero-mean fluctuations whose variance overwhelms the ensemble mean. We therefore restrict our analysis of the first-order coefficients to the theoretical predictions shown in \figref{fig:DTheoryFirstOrder}, with the corresponding scalings detailed in \appref{app:sec:ScalingRelation}.

\section{Scaling relation}\label{app:sec:ScalingRelation}
\begin{figure*}[!htbp]
	\centering
	\includegraphics[width =\textwidth]{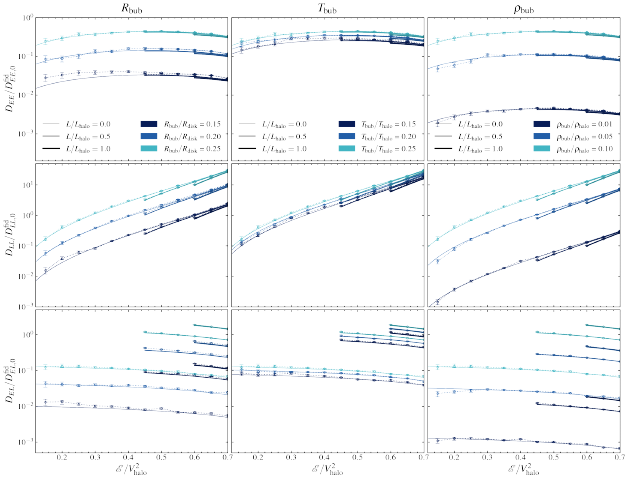}
	\caption{
	Measured diffusion coefficients $D_{EE}$, $D_{LL}$, and $D_{EL}$, scaled by corresponding fiducial diffusion coefficients, denoted by superscript ``fid'',  (from top to bottom panel) in \Mkv\@ simulation for different \SB\@ parameters $\Rbub$, $\Tbub$, and $\rhobub$ (from left to right panel) as a function of energy at a given angular momentum. Line contours are overlaid from the corresponding theoretical prediction. Color codes represent varying physical values in each panel, and the thickness of the lines shows angular momentum. 
	For non-radial orbits ($L/\Lhalo = 0.5$ and 1.0), the curves do not span the full energy axis, because no orbit with angular momentum $L$ can have an energy below that of the circular orbit with the same $L$. The shifted energy is therefore restricted to ${\calE/\Vhalo^{2} \in [\calE_{\mathrm{c}} / \Vhalo^{2} ,1]}$, where $\calE_{\mathrm{c}} = \Lc^{2}/(2 \rc^{2}) + \Phi(\rc)+\Vhalo^{2}$ is the shifted energy of the circular orbit of the radius $\rc(L)$ (see equation~\ref{eq:Lc}).
	Overall, our theoretical predictions are in good agreement with \Mkv\@ simulation for a range of parameters, validating the scaling laws given by equations~\eqref{eq:D0sets}. 
	}
	\label{fig:DiffusionCoefficientScaling}
\end{figure*}
Let us outline the derivation of approximate scaling relations introduced in \secref{subsec:ScalingRelations}, extracting three key quantities: (i)~the normalization amplitude $\Dz$, (ii)~the corresponding characteristic energy $\calEmax$ at maximum, and (iii)~the functional forms of diffusion coefficients. 

We first examine the normalization factor $\Dz$, whose exact amplitude depends strictly on the local properties of the perturbing sources, namely the spatial profile of the density (or potential) fluctuations. The density fluctuation is characterised by $\rhobub$ in equation~\eqref{eq:SBMassDensity}, and the corresponding potential fluctuation power spectrum in Fourier space scales as ${\tilde{\calP}_{\bk} \propto k^{-4} |\delta \tilde{\rho}_{\bk}|^{2}}$, following equations~\eqref{eq:LocalPowerSpectrum} and~\eqref{eq:PotentialToDensityPS}. Taylor expanding this power spectrum in the limit of small-sized, short-lived \SBs\@, where ${k \, \Rbub \ll 1}$ and ${\omega \, \Tbub \ll }1$, yields the approximate amplitude scaling ${\tilde{\calP}_{\bk} \propto (\dotnbubz \thalo/\Tbub) \, (\rhobub^{2} \, \Rbub^{5} \Tbub^{2}) \propto \dotnbubz \, \rhobub^{2}  \Rbub^{5} \Tbub }$, where the factor $\thalo/\Tbub$ accounts for the number of independent \SB\@ encounters per halo dynamical time. Consequently, the velocity diffusion coefficient scales as ${d_{\bv \bv} \propto \dotnbubz \, \rhobub^{2} \, \Rbub^{5} \Tbub}$. This parametric dependence inherently carries over to the diffusion in both energy and angular momentum. Qualitatively, this scaling stems from the spherical geometry of the \SBs\@, dictated by their characteristic size $\Rbub$, and their rapid ballistic decay, parameterised by the timescale $\Tbub$. The amplitude of the orbit-averaged diffusion coefficient in velocity retains the same scaling
\begin{align}
	\Dz &\!\approx\! \frac{512}{9375} \sqrt{\frac{14}{5}}\pi^{5/2}
	\left( \dotnbubz \, \Rhalo^{3} \right)
	\left(\frac{G \rhobub}{\thalo^{-2}} \right)^{2}
	\left(\frac{\Rbub}{\Rhalo}\right)^{5}
	\left( \frac{\Tbub }{\thalo} \right) \notag \\
	& = A \left( \dotnbubz \, \frac{4\pi}{3} \Rhalo^{3} \right) \left(\frac{\Ebub}{\Vhalo^{2}} \right)^{2}
	\left(\frac{\Rbub}{\Rhalo}\right)
	\left( \frac{\Tbub }{\thalo} \right)  \, ,
\end{align}
where the numerical prefactor $A$ obeys
\begin{equation}\label{eq:deconstant}
	A = \frac{72}{3125} \sqrt{\frac{14}{5 \pi}} \approx 0.02 \, .
\end{equation}
It reflects the fact that while \SBs\@ explode within the disc, only a subset of \SBs\@, which occupy the volume  $A 4\pi \Rhalo^3/3\approx 4\pi (0.275 \Rhalo)^3/3$, effectively disrupt the cusp. Equation~\eqref{eq:deconstant} is a non-trivial result of this work, in as much as the scalings in equation~\eqref{eq:D0sets} can be inferred by dimensional analysis, but the exact prefactor is the result of this Appendix.

The functional dependence on $(\calE, L)$ is dictated by the combination of the local diffusion coefficients in equations~\eqref{eq:LocalDiffusionCoefficientInEL} and the global spatial distribution of perturbing sources in equation~\eqref{eq:SigmabubInR}, as encoded within the orbit-averaging integrals. In particular, evaluating the analytical expressions for the local diffusion coefficients under the local resonance condition, $\omega = \bk \cdot \bv$, reveals the scaling of the integrand. For instance, for a uniform source distribution $\Sigmabub = 1$, the approximate diffusion coefficients in velocities follow the functional form
\begin{align}
\begin{bmatrix}
	d_{\vparl \vparl} \\ \addlinespace
	d_{\vperp \vperp} \\ \addlinespace
\end{bmatrix}
&=  \frac{3}{2} \Dz \Vhalo^{2}
\begin{bmatrix}
(1 + B v^{2})^{-3/2} \\ \addlinespace
 \tfrac{\pi}{2} (1 + B v^{2})^{-1/2} \\ \addlinespace
 \end{bmatrix}\,,
\end{align}
where $B = 7\Tbub^{2}/(100 \pi \Rbub^{2})$. The corresponding local diffusion coefficients in energy and angular momentum are then
\begin{align}
\begin{bmatrix}
d_{EE} \\ \addlinespace
d_{LL} \\ \addlinespace 
d_{EL}
\end{bmatrix}
&=\frac{3}{2} \Dz \Vhalo^{2}
\begin{bmatrix}
v^{2} \\ \addlinespace
\tfrac{L^{2}}{v^{2}} + \tfrac{1}{\pi} \left(r^{2} - \tfrac{L^{2}}{v^{2}} \right) (1 + B v^{2})\\ \addlinespace
L
\end{bmatrix} \, .
\end{align}
Assuming a large disc scale radius (${r/\Rdisc \ll 1}$) for the inhomogeneous distribution given by equation~\eqref{eq:SigmabubInR}, together with the spatial locality of the \SBs\@, the functional form of the orbit-averaged diffusion coefficients can be approximated as multiplicative combinations of $\calE$, $1-\calE$, and $L$ raised to appropriate powers, with the location of the maximum specified by equation~\eqref{eq:EMaxApproxPade}. Note that these powers vary according to the functional dependence of the integrand on $r$ and $v$, which is governed by the spatial distribution and geometry of the perturbers.

Despite the approximations based on dimensional matching and local perturbations, the resulting expressions agree well with exact predictions (e.g., \figref{fig:DTheory3D}) and are validated by numerical simulations (e.g., \figref{fig:DMarkovian3D}). In part, this agreement holds because incoherent orbit-averaging does not drastically alter the functional dependence of the orbit-averaged diffusion coefficients compared to their local counterparts. 

In addition to the second-order diffusion coefficients and their corresponding approximate scalings presented in \secref{subsec:GeneralOrbits}--\secref{subsec:ScalingRelations}, the first-order diffusion coefficients can be derived in an analogous manner, as illustrated in \figref{fig:DTheoryFirstOrder}. This yields the approximate scaling relation:
\begin{subequations}\label{eq:ScalingRelationFirstOrder}
\begin{align}
	D_{E}^{\text{approx}} &\!=\! D_{E, 0}  \,  \left( 1 - \frac{\calE}{2 \calEmax} \right)^{2} \left(1- \frac{L}{8\Lc[\calE]} \right)^{-3/4}  \, , \label{eq:DEApprox}\\
	D_{L}^{\text{approx}} &\!= \! D_{L,  0}  \left( 1 - \frac{\calE}{2 \calEmax} \right)^{\! -1} \left(  \frac{L}{8\Lc[\calE]}\right)^{\! -1} \left(1- \frac{L}{8\Lc[\calE]} \right)^{\! -3/4} \!\!\! \!\!\! , \label{eq:DLApprox}
\end{align}
\end{subequations}
where the corresponding normalization factors, following equations~\eqref{eq:ScalingRelation}, are
\begin{align}\label{eq:D0sets2}
	\begin{bmatrix}
	D_{E, 0} \\
	D_{L, 0}
	\end{bmatrix}
	&= D_{0}
	\begin{bmatrix}
	\Vhalo^{2} \\
	5 \Rhalo \Vhalo
	\end{bmatrix} \, .
\end{align}

\section{Validity of \Mkv\@ assumption}\label{app:sec:ValidityOfMarkovianAssumption}
\begin{figure}[!htbp]
	\centering\includegraphics[width=0.42\textwidth]{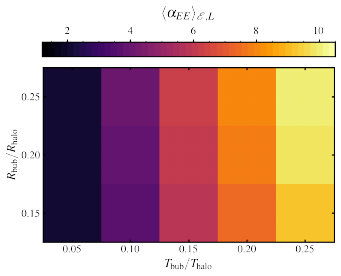}
	\caption{
	Averaged ratio $\langle \alpha_{EE} \rangle_{\calE, L}$ of the diffusion coefficient in energy measured in \nMkv\@ and \Mkv\@ simulations over all $(\calE, L)$ bins, 
	denoted by angle brackets with subscripts. The corresponding average, $\langle \alpha_{LL} \rangle_{\calE, L}$ and $\langle \alpha_{EL} \rangle_{\calE, L}$, exhibit very similar trends to $\langle \alpha_{EE} \rangle_{\calE, L}$, within differences of less than $3\%$ in their respective values. 
	 As the \SB\@ perturbations become more localized in space and time, $\alpha_{EE}$ approaches unity, where the \Mkv\@ assumption captures accurately the amplitude of diffusion measured in \nMkv\@ simulation. The distinct scaling behaviours of $\alpha$ with $\Rbub$ and $\Tbub$ further imply that the spatial and temporal correlations have different characteristic extents.
	 }	
	 \label{fig:NonMarkovianConvergence2}
\end{figure}

Although the orbit-averaging formalisms based on the \Mkv\@ assumption between successive scatterings have successfully predicted the secular evolution of dynamically hot systems \citep[e.g., ][]{Theuns1996,Fouvry2021,Tep2022,Tep2024,Tep2026}, discrepancies with numerical simulations can still persist because these formalisms neglect correlations between successive encounters as well as the effects of strong, non-perturbative interactions. To assess their impact beyond the perturbative regime examined in \secref{sec:NumericalValidation}, let us examine the validity of the \Mkv\@ assumption. 

To quantify departures from \Mkv\@ behaviour, let us define the ratio of diffusion coefficients measured in the \nMkv\@ and \Mkv\@ simulations as ${\alpha_{XY} \equiv D^{\text{NM}}_{XY} / D^{\text{M}}_{XY}}$, where the subscripts $XY$ denote the corresponding combination of $E$ and $L$. We compare $\alpha_{EE}$, $\alpha_{LL}$, and $\alpha_{EL}$ as the characteristic size, lifetime, and density of \SBs\@ are varied from the fiducial values. Overall, the ratio converges toward unity, homogeneously across most of the $(\calE, L)$ phase space, except in the limit $\calE \to 0$ and $L \to 0$,\footnote{These regimes correspond to tightly bound and nearly radial orbits, respectively, and may be especially sensitive to the spatial and temporal structure of the perturbations (e.g., overlapping \SBs\@ or breakdown of the independent-encounter approximation).
} as the perturbers become more localized (e.g., decrease in $\Rbub$ and $\Tbub$) or as their amplitude is reduced (e.g., decrease in $\rhobub$). 

To illustrate this behaviour, \figref{fig:NonMarkovianConvergence2} displays $\alpha_{EE}$ averaged over all $\calE, L$ bins as a function of $\Rbub/\Rhalo$ and $\Tbub / \thalo$ for a fixed density ratio $\rhobub / \rhohalo = 0.1$.\footnote{In practice, we probe the parameter space down to $\Rbub / \Rhalo = 0.15$ and $\Tbub / \thalo = 0.05$, as limited numerical resolution can otherwise underestimate the diffusion coefficients (see \appref{app:secTruncationInFrequency}). Moreover, because the diffusion timescale follows the power-law scaling of equation~\eqref{eq:tD}, probing smaller scales requires progressively longer integration times to reliably measure slopes within the diffusive regime.} These trends suggest that the validity of the \Mkv\@ approximation depends on the spatial and temporal scales of the perturbations over the parameter range considered. This departure from \Mkv\@ behaviour can be understood in terms of the distinct roles played by the perturbation parameters in violating local \Mkv\@ assumptions.

The characteristic size $\Rbub$ and lifetime $\Tbub$ directly determine the spatial and temporal correlations. Assuming uniformly random birth locations and times, smaller and shorter-lived perturbers provide greater spatial and temporal localization, thereby weakening correlations between successive encounters and improving the applicability of the \Mkv\@ assumption.

In contrast, $\rhobub$ primarily controls the perturbation's amplitude. Because the fiducial \SB\@ mass density is already an order of magnitude below the halo density, its quadratic scaling in the power spectrum exerts a minor effect on the normalized correlation structure. Consequently, within the probed density range, decreasing $\rhobub$ does not significantly alter the departure from \Mkv\@ behaviour compared to variations in $\Rbub$ and $\Tbub$, whereas a more pronounced departure is expected for larger values of $\rhobub$.
A comparison of the spatial and temporal variations between $\Rbub$ and $\Tbub$ suggests that the rate of convergence towards the \Mkv\@ regime seems more pronounced when $\Tbub$ decreases than when $\Rbub$ decreases by a comparable relative amount. This difference indicates that temporal and spatial correlations  have distinct characteristic scalings, encoded in their  contribution to the local power spectrum, given by equation~\eqref{eq:PotentialToDensityPS}.

\section{Evolution of density profile}\label{app:sec:EvolutionOfDensityProfile}
Once the orbit-averaged diffusion coefficients are specified as described in \secref{sec:Results}, we have access to the time evolution of the \DF\@, by solving the \FP\@ equation in equation~\eqref{eq:FokkerPlanckInEL}. In the context of cusp-to-core transformations, this allows us to predict the time-dependent mass density profile. 

To characterise the flattening of the central cusp as a function of time, we consider the generalized $\gamma$-model density profile \citep{Hernquist1990,Zhao1996}, extended from the original \NFW\@ profile given by equation~\eqref{eq:NFWDensity}, which follows
\begin{equation}\label{eq:GeneralizedNFWDensity}
	\rho(r) = \frac{\rhohalo}{\left( r/\Rhalo  \right)^{\gamma} {\left[1+( r/\Rhalo)^{\mu}  \right]^{(\nu - \gamma) / \mu}}} \, ,
\end{equation}
where $\gamma$ (resp. $\nu$) controls the slope of the inner (resp. outer) profile and $\mu$ the sharpness of the transition. Note that equation~\eqref{eq:GeneralizedNFWDensity} reduces to equation~\eqref{eq:NFWDensity} for $(\mu, \nu, \gamma) = (1, 3, 1)$. In this parameterization, tracking the cusp-to-core transformation can be characterised by a decrease in $\gamma$ as a function of time. 

To reconstruct the mass density profile from the evolving \DF\@, we use  Eddington's formula (e.g., \citealt{Eddington1916}; \S{4.3.1} from \citetalias{BinneyTremaine2008}). For an isotropic system with $f = \fiso(E)$, the density reads 
\begin{equation}\label{eq:EddingtonFormula}
	\varrho = 4 \sqrt{2} \pi \int_{0}^{\psi} \! \dd \xi \, (\psi - \xi)^{1/2} \, \fiso(\xi) \, .
\end{equation}

As the mean-field halo potential is assumed to be dominant over the perturbations, we treat the background potential as fixed when evolving the \DF\@. Thus, for simplicity, we reconstruct the time-dependent density profile by substituting the solution to the \FP\@ equation into equation~\eqref{eq:EddingtonFormula}, instead of iteratively updating the potential in response to the changing density \citep{Weinberg2001}.

Recall the \FP\@ equation with the approximate diffusion coefficient in energy $D_{EE}^{\text{approx}}$ given by equation~\eqref{eq:DEEofEScaling}. By introducing the rescaled time $\tau \equiv t / \tdiff$ with diffusion timescale defined in equation~\eqref{eq:tD}, the governing equation can be rewritten as
\begin{align}\label{eq:ApproximateFPEquation}
	\frac{\p f}{\p \tau} = 2 \Vhalo^{4} \frac{\Vhalo^{2}}{\calEmax} \frac{\p}{\p \calE} 
	\left[ 
	\left( \frac{\calE}{\Vhalo^{2}} \right)
	\left(1 - \frac{\calE}{2 \calEmax} \right)
	\frac{\p f}{\p \calE} 
	\right] \, .
\end{align}

We explore two approaches to solving equation~\eqref{eq:ApproximateFPEquation}, distinguished by their numerical treatments and boundary conditions: (i)~a spectral method that does not explicitly enforce mass conservation, and (ii)~a zero-flux numerical scheme. In both approaches, we restrict the energy domain to $\calE \in [0, 2\calEmax]$ to ensure the non-negativity of the quadratic second-order approximate diffusion coefficients.

For the spectral method, we seek a time-dependent solution via an eigenfunction expansion of the form
\begin{align}\label{eq:fisoSol}
	f(\calE, \tau) = \sum_{n} \phi_{n}(\calE) \, \mathrm{e}^{-\lambda_{n} \tau} \, ,
\end{align}
where $\phi_{n}(\calE)$ is the eigenfunction associated with the eigenvalue $\lambda_{n}$. Because the forward integral in the Eddington inversion formula is linear in $f$, substituting equation~\eqref{eq:fisoSol} into equation~\eqref{eq:EddingtonFormula} directly yields the temporal evolution of the density profile. In this scheme, the density of states associated with the \DF\@ is held fixed, so total mass conservation is not explicitly enforced.

In the second approach, we solve equation~\eqref{eq:ApproximateFPEquation} numerically by imposing zero-flux boundary conditions at $\calE = 0$ and $\calE = 2\calEmax$. These conditions prevent probability flow across the boundaries and guarantee the conservation of total integrated quantities, provided the \FP\@ equation is written in conservative form.

To isolate the evolution of the inner logarithmic density slope $\gamma$, we assume the initial density profile follows equation~\eqref{eq:GeneralizedNFWDensity} with parameters ${(\mu_{0}, \nu_{0}, \gamma_{0}) = (1, 3, 1)}$ at ${\tau = 0}$, and that only $\gamma(\tau)$ varies dynamically. Thus, in the inner region ${r / \Rhalo \ll 1}$, where the density profile scales as ${\rho(r, \tau) \propto r^{-\gamma(\tau)}}$, taking the logarithmic time derivative yields ${\p \ln \rho / \p \tau = -\dot{\gamma} \ln r}$. Evaluating this rate at early times ($\tau \ll 1$) gives $(\mathrm{d} \gamma / \mathrm{d} \tau) / \gamma = -1$, which upon restoring physical time units recovers the scaling relation in equation~\eqref{eq:GammaScalingIsotropicDF}.

Although this approach neglects the back-reaction of the mean-field by isolating diffusive effects, this assumption is consistent with the incoherent orbit-averaging framework established in \secref{sec:Theory}, which has been shown to hold in dynamically hot systems \citep[e.g.,][]{Fouvry2021,Tep2022,Tep2024,Tep2026}. A fully self-consistent treatment can also be rendered tractable numerically \citep[e.g.,][]{Vasiliev2017}.

\figref{fig:GammaEvolution} shows the evolution of $\gamma$ as a function of time.  We observe a good match between the various approximations, all displaying the expected cusp-to-core transformation. 

\section{Truncation in frequency}\label{app:secTruncationInFrequency}
\begin{figure}[!htbp]
	\center \includegraphics[width=0.49\textwidth]{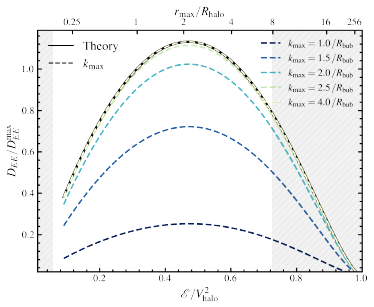}
	\caption{
	$D_{EE}$ from the theoretical prediction for radial orbits (black solid line), and that with different cut-off frequency scale $\kmax$ in the unit of $1/\Rbub$ (color coded dashed lines). For comparison, the $x$-axis on the top is given by the maximum semi-major axis $\rmax$ at a given $\calE$. The diffusion coefficients with insufficient resolution (e.g., lower $\kmax$) tend to be underestimated, compared with considering the full power spectrum range ($\kmax \to \infty$). 
	}
	\label{fig:DEEkmax}
\end{figure}
Let us consider the impact of resolution on the measurement of the diffusion coefficient. Ideally, one should account for the full frequency range of the power spectrum; in practice, however, finite resolution might introduce cut-offs. Such cut-offs may limit the accessible ranges of scales and lead to over- or under-estimation of the observables by disregarding dynamical contributions beyond the resolution limit. 

Motivated by this, we consider the maximum spatial frequency scale $\kmax$ in equation~\eqref{eq:LocalDiffusionCoefficientInVelocity}, modified as
\begin{equation}\label{eq:LocalDiffusionCoefficientInVelocityKmax}
 d_{v_{i} v_{j}}^{\kmax} (\bv) = \dotnbub \int_{[-\kmax, \kmax]^{3}} \! \frac{\dd^{3} \bk}{(2\pi)^{3}} \, k_{i} k_{j} \, \tilde{\calP}_{\bk}(\omega = \bk \cdot \bv) \, ,
\end{equation}
and measure the $\DEErad$. The results with different values of $\kmax$, compared to the un-truncated result, are illustrated in \figref{fig:DEEkmax}. 

Overall, when $\kmax$ is insufficient to resolve scales smaller than approximately half the \SB\@ radius, diffusion is underestimated. However, this discrepancy is progressively eliminated as $\kmax$ increases. By performing a Taylor expansion in the large-${\kmax}$ limit, we obtain the following scaling relation 
\begin{equation}\label{eq:DEEofEScalingKmax}
	D_{EE}^{\kmax} (\calE) \approx 
	D_{EE} (\calE)
	\left(1 - \frac{5}{14 \, \kmax \Rbub} \right) \, .
\end{equation}
We highlight that the impact of resolving small scales on global diffusion is not specific to a particular power spectrum. As discussed in \secref{subsec:LocalAveraging}, the subset of all scales that satisfy the resonance condition will contribute to the local diffusion coefficient, with the relative contribution of each scale depending on the properties of the perturbers and corresponding power spectrum. 

For example, \citet{ElZant2016} adopted a power-law power spectrum, and this specific scale-free set-up by design makes the normalization factor and the largest scales drive the diffusion most dominantly. While this provides a useful illustration of the large-scale fluctuations, its result does not extend to general power spectra from realistic perturbations in galactic systems.

Even when incorporating more sophisticated power spectra, selecting an appropriate wavenumber or frequency range remains essential when comparing theoretical predictions with numerical measurements. If a simulation or measurement setup fails to capture small-scale fluctuations below its resolution limit (e.g., via  specific choice of sub-grid physics), or large-scale modes exceeding the probed domain ({e.g.,} while using finite simulation box size), it risks underestimating the cumulative impact of these fluctuations on long-term secular evolution.

\section{Superbubble model parameters}\label{app:sec:Parameters}
In \appref{app:subsec:Superbubble} and \appref{app:subsec:Galaxy}, we respectively summarize the fiducial physical parameters of the \SBs\@, \DM\@ halo, and stellar disc used in the numerical simulations of \secref{sec:NumericalValidation} and the diffusion timescale estimates of \secref{subsec:Galaxies}.
\subsection{Superbubble parameters}\label{app:subsec:Superbubble}
Bubble-like structures originate from diverse formation mechanisms, including pre-supernova feedback and photoionization \citep[e.g.,][]{Lim2020}, stellar winds from OB associations \citep[e.g.,][]{OstrikerMcKee1988}, and multiple clustered associations \citep[e.g.,][]{Nath2020}, thus spanning several orders of magnitude in spatial size and lifespan. Although \SBs\@ are primarily associated with the last mechanism, their continuous distribution across parameter space precludes defining a single representative set of parameters for a \SB\@. Within the scope of this paper, we therefore provide physical values not as definitive measurements, but as order-of-magnitude estimates to guide physical interpretation. We summarize a representative set of observational \SB\@ parameters in Table~\ref{tab:SBParametersFromObservation}.
\begin{table}[!h]	
	\caption{Physical \SB\@ parameters adopted in equation~\eqref{eq:tDwithNumbers}}
	\label{tab:SBParametersFromObservation}
	\centering
	\begin{threeparttable}
		\small
		\setlength{\defaultaddspace}{0.25em}
		\begin{tabular*}{0.9\columnwidth}{@{\extracolsep{\fill}}lc}
			\toprule
			Parameter $[\mathrm{unit}]$ & Value \\
			\midrule
			$\Rbub \, [\pc]$\tnote{a} & 250 \\ \addlinespace
			$\Tbub \, [\Myr]$\tnote{b} & 25   \\ \addlinespace
			$\rhobub \, [\Msol \, \pc^{-3}]$\tnote{c} & 0.05 \\ \addlinespace
			$\dotnbubz \, \Rhalo^{3} [\Gyr^{-1}]$\tnote{d} & $7.3 \times 10^{-3} \left( \dfrac{\overline{\SFR} / [\Msol \, \Myr^{-1}]}{\overline{M}_{\text{sc}} / [1.4 \times 10^{5} \, \Msol]} \right) \left(\dfrac{\Rdisc}{\Rhalo} \right)^{3}$  \\
			\bottomrule
		\end{tabular*}
		\begin{tablenotes}
			\scriptsize
			\item[a] Inner radius of \SB\@. Outer radius is set to be twice this value (see \appref{app:subsec:EvolutoinOfSuperbubbles}).
			\item[b] Lifetime of \SB\@.
			\item[c] Inner mass density of \SB\@. The outer mass density  yields zero net perturbed mass.
			\item[d] Rate of encountered \SBs\@, derived from the mean \SFR\@ at a given halo mass, $\overline{\SFR}$ (see Table~\ref{tab:GalaxyParameters}), and mean star cluster mass that is expected to produce one \SB\@ ${\overline{M}_{\text{sc}}}$ (see \citealt{Yadav2017}; \S{2} in \citealt{Kipper2025}).
		\end{tablenotes}
	\end{threeparttable}
\end{table}

For the \SB\@ radius, lifetime, and mass density, we adopt ${\Rbub = 250 \, \pc}$, ${\Tbub = 25 \, \Myr}$, ${\rhobub = 0.05 \, \Msol \, \pc^{-3}}$, and ${\dotNbub = 1.0}$, based on previous works in \SB\@ hydrodynamical simulation \citep{Yadav2017,Nath2020,Kipper2025}. Note that this choice corresponds to larger \SB\@ structures, but still within the representative regimes \citep{Watkins2023}.  

Inevitably, this parameter set carries intrinsic uncertainties: the difficulty of establishing definitive values stems from non-uniform measurement methods, inconsistent definitions across studies, and resolution limits \citep{Watkins2023}. Beyond these observational challenges, the \SB\@ parameters are partially degenerate, as their temporal evolution scales jointly with \SB\@ size and  density.
Despite these uncertainties, the kinetic framework remains robust because the \SB\@ parameters are explicitly factored out into a single overall normalization factor $D_{0}$ in equation~\eqref{eq:D0sets}, allowing observational uncertainties and parameter degeneracies to be absorbed directly into this prefactor. If needed, one could account for a continuous distribution of perturber scales $(\Rbub, \Tbub)$ by performing an additional weighted average over the diffusion coefficients.

For the numerical simulation in \secref{sec:NumericalValidation}, we adopt the fiducial \SB\@ parameters as ${\Rbub/\Rhalo = 0.25}$, ${\Tbub/\thalo = 0.25}$, and ${\rhobub/\rhohalo = 0.1}$. For dwarf galaxies with halo masses ${10^{8\text{--}9} \, \Msol}$, these ratios broadly reproduce the physical values in Table~\ref{tab:SBParametersFromObservation}. For more massive halos, the physical values yield smaller ratios, placing the mean-field system deeper in the perturbative regime, where our framework is well validated (see \appref{app:sec:ValidityOfMarkovianAssumption}).

As an occasional outburst of the ambient medium, we approximate \AGN\@ feedback-driven perturbation as a single \SB\@ at the galactic centre \citep[e.g.,][]{Weinberg2025}. Although \AGN\@ bubbles are buoyant, larger and longer-lived than \SBs\@, we assume their dynamical structure remains similar in the perturbative, short relative velocity regime. The main difference is the driving force: a fraction ${\epsilon\!\sim 0.1}$ of the rest-mass energy of gas accreted onto the central black hole, ${E_{\text{AGN}} =  \epsilon \,\Delta M_{\text{acc}} c^{2}}$, where ${\Delta M_{\text{acc}}}$ is the accreted gas mass and $c$ is speed of light, pushes the \AGN\@ bubble by displacing ambient gas of mass ${\Delta M_{\text{gas}}}$ against a potential of circular velocity $\vc$. This yields ${E_{\text{AGN}} \sim \Delta M_{\text{gas}} \vc^2}$ with displaced gas mass ${\Delta M_{\text{gas}}}$. For a ${10^{14} \, \Msol}$ halo, we adopt in equation~\eqref{eq:tDwithAGN} the large-bubble values ${R_{\text{AGN}} = 30 \, \kpc}$, and ${T_{\text{AGN}} = 100 \,\Myr}$, and  ${\rho_{\text{AGN}} \simeq 3\times10^{-4}\,\Msol\,\pc^{-3}}$,
giving ${\Delta M_{\text{gas}} \simeq 3\times10^{10}\,\Msol}$ and, for ${\vc \simeq 650\,\kmPs}$, ${E_{\text{AGN}} \sim 3\times10^{59}\,\erg}$.

This is typical of cavity energetics in groups and clusters \citep{Birzan2004,Rafferty2006}: only ${\sim10^{-4}}$ of ${\Mbh c^{2}}$ for the central black hole mass ${\Mbh \sim 10^{9}\,\Msol}$. Again, these values should be considered as reference model parameters rather than accurate estimates, since observed \AGN\@ bubbles span a broad range. Depending on the host galaxy and outburst from central black holes, ${(R_{\text{AGN}} \, [\kpc], T_{\text{AGN}} \, [\Myr],\rho_{\text{AGN}} \, [\Msol \, \pc^{-3}])}$ ranges from ${(\sim\!\!5, \sim\!\!5, \sim\!\!10^{-5} )}$ for the Fermi bubbles~\citep{MillerBregman2016}, through ${(\sim\!\!7, \sim\!\!20, \sim\!\!10^{-5})}$ for the eROSITA bubbles \citep{Predehl2020}, to ${(\gtrsim\!\!100,\sim\!\!100, \sim\!\!10^{-4})}$ for the largest bubbles \citep{McNamaraNulsen2007}.

\subsection{Galaxy parameters}\label{app:subsec:Galaxy}
The \DM\@ halo \NFW\@ profile is determined by two parameters $(\Rhalo, \Vhalo)$ or $(\Rhalo, \rhohalo)$. In practice, typical observables are given by the virial mass $\Mvir$, virial radius $\Rvir$ and concentration parameter $\cNFW$, defined as
\begin{align}\label{eq:VirialMassDefinition}
	\Mvir= \frac{4\pi}{3} \Rvir^{3} \times 200 \, \rhocrit \, ; & \quad{} \cNFW = \frac{R_{200}}{\Rhalo} \,,
\end{align}
where $\rhocrit= 3 H^{2}/ (8 \pi G) $ is the critical density. For \NFW\@ halo, integrating equation~\eqref{eq:NFWDensity} from 0 to $\Rvir$ gives the virial mass
\begin{align}\label{eq:VirialMass}
	\Mvir &= \int_{0}^{\Rvir} \dd^{3} \br \frac{\rhohalo}{\left( r/\Rhalo \right) \left(1 +  r/\Rhalo \right)^{2}} \notag \\
	&= 4 \pi \rhohalo \Rhalo^{3} \left[\log(1+\cNFW) - \frac{\cNFW}{1+\cNFW} \right] \, .
\end{align}
Equating equation~\eqref{eq:VirialMass} to \eqref{eq:VirialMassDefinition} gives the conversion from observables to halo parameters, which reads
\begin{equation}
	\rhohalo = \frac{ 200 \rhocrit }{3} \frac{\cNFW^{3}}{\left[ \log(1+\cNFW) - \cNFW/(1+\cNFW) \right]} \, ,
\end{equation}
and
\begin{equation}
	\Rhalo = \frac{1}{\cNFW} \left(\frac{3}{4\pi} \frac{\Mvir}{200 \rhocrit } \right)^{1/3} \, .
\end{equation}
The physical values for \DM\@ halo and stellar disc parameters are summarized in Table~\ref{tab:GalaxyParameters}.
\begin{table}[!h]	
	\centering
	\caption{
	Physical \DM\@ halo and stellar disc parameters for typical dwarf, \MW\@, and giant galaxies.
	}
	\label{tab:GalaxyParameters}
	\begin{threeparttable}
	\small
	\setlength{\defaultaddspace}{0.25em}
	\begin{tabular*}{0.9\columnwidth}{@{\extracolsep{\fill}}lccc}
	\toprule
	 Parameter $[\text{unit}]$ & dwarf & \MW & giant \\
	\midrule
	$\Mhalo \, [\Msol]$\tnote{a} & $10^{10}$ & $ 10^{12}$ & $10^{14}$ \\ \addlinespace
	$\cNFW \, [\cdot]$\tnote{b} & $ 13.3$ & $8.4$ & $ 5.3$ \\ \addlinespace
	$\rhohalo \, [\Msol \, \pc^{-3}]$\tnote{c}  & $1.1 \times 10^{-2}$  & $3.6 \times 10^{-3}$ & $1.2 \times 10^{-3}$ \\ \addlinespace
	$\Rhalo \, [\kpc]$\tnote{d} & 3.4 & 25.4 & 187.4 \\ \addlinespace
	$\thalo \, [\Myr]$\tnote{e} & 39.4 & 69.8 & 120.5 \\ \addlinespace	
	$\Vhalo \, [\kmPs]$\tnote{f} & 85.1 & 355.2 & 1520.0 \\ \addlinespace
	$\Rdisc \, [\kpc]$\tnote{g} & 9.1 & 42.4 & 197 \\ \addlinespace
	$\overline{\SFR} \, [\Msol \, \Myr^{-1}]$\tnote{h} & $10^{4}$ & $2 \times 10^{6}$ &  $1 \times 10^{6}$ \\
	\bottomrule
	\end{tabular*}
	\begin{tablenotes}
		\scriptsize
		\item[a] \NFW\@ \DM\@ halo mass enclosed within $R_{200}$, given by equation~\eqref{eq:VirialMass}.
		\item[b] Concentration parameter defined by equation~\eqref{eq:VirialMassDefinition}. The \NFW\@ $M_{200}-\cNFW$ relation at $z = 0$ from equation~(8) in \cite{Dutton2014} was adopted.
		\item[c] \NFW\@ \DM\@ halo mass density, given by equation~\eqref{eq:NFWDensity}.
		\item[d] \NFW\@ \DM\@ halo scale radius.
		\item[e] \NFW\@ \DM\@ halo timescale.
		\item[f] \NFW\@ \DM\@ halo characteristic velocity.
		\item[g] Stellar disc scale radius defined by equation~\eqref{eq:SigmabubInR}.
		\item[h] \SFR\@ of galaxies, which determines the number of \SBs\@ in $\dotNbub$ (see Table~\ref{tab:SBParametersFromObservation}). 
	\end{tablenotes}
	\end{threeparttable}
\end{table}

We adopt the halo mass--concentration relation at redshift $z=0$ derived from 
N-body simulations in the Planck cosmology (see equation~8 of \citealt{Dutton2014}).\footnote{
At higher redshifts, the concentration at a fixed halo mass is lower (see equations~10--11 of \citealt{Dutton2014}). For fixed fiducial \SB\@ parameters, this increases the relative \SB\@-to-\DM\@ ratio and thus shortens the overall diffusion timescale. In practice, however, progenitor haloes at higher redshifts are also less massive and exhibit lower \SFRs\@, which reduces the \SB\@ amplitudes and partially compensates for this effect. We therefore neglect redshift evolution here, as we expect it to alter the amplitude of the diffusion coefficients by at most a factor of a few over the timescales of interest, while the system remains mean-field dominated.
} 
The \NFW\@ parameter values listed in Table~\ref{tab:GalaxyParameters} are broadly consistent with observations of dwarfs \citep{Read2017}, the \MW\@ \citep{BlandHawthorn2016,SylosLabini2023}, and giant galaxies \citep{Ettori2010}.

For the disc scale radii in equation~\eqref{eq:SigmabubInR}, we adopt the relation ${\Rdisc \sim 0.2 \, R_{200}}$, motivated by the relation between the stellar half-mass radius and halo virial radius found empirically \citep{Kravtsov2013} and predicted by angular-momentum model \citep{Mo1998}. Our choice of $0.2$ is, however, 5--20 times larger than these works, and the resulting \MW\@ value is about eight times the measured scale length of $\simeq 2.6 \, \kpc$ using the exponential disc profile \citep{Licquia2015,BlandHawthorn2016}. Nevertheless, we make this choice deliberately: the approximations used to derive the scaling relations rely on orbits within $\Rdisc$ (see \appref{app:sec:ScalingRelation}), and a larger $\Rdisc$ keeps the $\calEmax$ of equation~\eqref{eq:EMaxApproxPade} within the range 0--1. Note that the exact theoretical calculation does not depend on this choice: the amplitude and shape of the diffusion coefficients change by less than a factor of two (see solid lines in \figref{fig:DEEofE}), and our conclusions are unaffected. The profile in equation~\eqref{eq:SigmabubInR} and the physical values of $\Rdisc$ should therefore be regarded as model parameters rather than as a faithful reproduction of observed galaxies.

Following \citet{Kipper2025}, we assume that each \SB\@ is formed within a single star cluster (subscript ``sc'' with overline) of the mean mass ${\overline{M}_{\text{sc}} \simeq1.4\times10^{5}\,\Msol}$, so that ${\dotNbub \simeq \overline{\SFR} / \overline{M}_{\text{sc}}}$, where the $\overline{\SFR}$ represents the mean \SFR\@ at a given halo mass, taken from the literature (dwarfs: \citealt{ReadErkal2019}; \MW\@: \citealt{Licquia2015,Read2017}; central galaxies of massive halos: \citealt{Behroozi2013,McDonald2018,OrellanaGonzalez2022}).

\clearpage
\end{CJK*}
\end{document}